\documentclass[12pt]{article} 
\usepackage{epsfig,graphicx,psfrag,axodraw}
\usepackage{amsmath,amsfonts,stmaryrd,latexsym}
\usepackage{hyperref}
\hypersetup{breaklinks=true, pagecolor=white, colorlinks=false}
\newcommand{\be}{\begin{equation}}
\newcommand{\ee}{\end{equation}}
\newcommand{\bea}{\begin{eqnarray}}
\newcommand{\eea}{\end{eqnarray}}
\newcommand{\bear}{\begin{eqnarray}}
\newcommand{\eear}{\end{eqnarray}}

\newcommand{\ba}{\begin{array}}
\newcommand{\ea}{\end{array}}
\newcommand{\lae}{\begin{array}{c}\,\sim\vspace{-21pt}\\<
\end{array}}
\newcommand{\gae}{\begin{array}{c}\,\sim\vspace{-21pt}\\>
\end{array}}
\newcommand{\Tr}{{\rm Tr}}

\begin{document}

\baselineskip=18pt \pagestyle{plain} \setcounter{page}{1}

\vspace*{-0.8cm}

\noindent 
\makebox[11.5cm][l]{\small \hspace*{-.2cm} December 28, 2010}{\small FERMILAB-Pub-10-534-T}  \\
\makebox[11.5cm][l]{\small \hspace*{-.2cm} }{\small SLAC-Pub-14343} \\ [-2mm]

\begin{center}

{\Large \bf Heavy octets and Tevatron signals with \\ [3mm]  three or four $b$ jets }\\ [9mm]

{\normalsize \bf Yang Bai$^1$ and Bogdan A. Dobrescu$^2$  \\ [4mm]
{\small {\it
$^1$ Theoretical Physics Department, SLAC, Menlo Park, CA 94025, USA \\ http://www.slac.stanford.edu/th \\ [2mm]
$^2$ Theoretical Physics Department, Fermilab, Batavia, IL 60510, USA \\ http://theory.fnal.gov }}\\
}

\end{center}

\vspace*{0.4cm}

\begin{abstract}
Hypothetical color-octet particles of spin 0, pair-produced at hadron colliders through their QCD coupling,
may lead to final states involving three or four $b$ jets.   
We analyze kinematic distributions of the $3b$ final state that differentiate the scalar octets from 
supersymmetric Higgs bosons. Studying the scalar sector that
breaks an $SU(3)\! \times\! SU(3)$ gauge symmetry down to the QCD gauge group, we find 
that the scalar octet is resonantly produced in pairs via a spin-1 octet (coloron).
A scalar octet of mass in the 140 -- 150 GeV range can explain the nonstandard shape of the $b$-jet transverse 
energy distributions reported by the CDF Collaboration, especially when the coloron mass is 
slightly above twice the scalar mass. 
The dominant decay mode of the scalar octet is into a pair of gluons, 
so that the production of a pair of dijet resonances is large in this 
model, of about 40 pb at the Tevatron. Even when a $W$ boson is radiated from the initial state, the inclusive 
cross section for producing a dijet resonance near the scalar octet mass remains sizable, around 0.15 pb.
\end{abstract}

\tableofcontents

\section{Pairs of resonances} \setcounter{equation}{0}

Hadron colliders, such as the Tevatron and the LHC, allow the production at a high rate of new 
particles carrying QCD color. The subsequent decays of such particles often involve only QCD jets,
which are hard to separate from the background. However, when the jets originate from $b$-quark decays,
the backgrounds are substantially reduced. Among the hypothetical particles leading to signatures 
involving several $b$ jets are color-octet bosons. The simplest of those is the weak-singlet particle
of spin 0, generically referred to here as the $G_H$ scalar, which is present in various theories 
including Technicolor (``octet technipion'' \cite{Hill:2002ap}), the 6D standard model (the Kaluza-Klein modes 
of the gluon polarized along extra dimensions \cite{Burdman:2006gy}), vectorlike confinement \cite{Kilic:2008pm}, 
weakly-interacting metastable pion models \cite{Bai:2010qg}, or certain supersymmetric models (``sgluons" \cite{Plehn:2008ae}). 
The $G_H$ scalar can be produced in pairs through 
its QCD couplings to gluons, and may decay through higher-dimensional operators into a pair of heavy quarks.
The signature is four $b$ jets, forming two $b\bar{b}$ resonances of same mass \cite{Dobrescu:2007yp, Chivukula:1991zk}.

Here we point out that an alternative way of searching for scalar octets is to require only three jets to pass the 
basic cuts and to be $b$ tagged. We show that in this case the shapes of certain kinematic distributions may 
allow the separation of the signal from the background, and the differentiation between various extensions of the standard model.
For example, compared to the Minimal Supersymmetric Standard Model (MSSM) at large $\tan\beta$, 
where a Higgs boson is produced in association with  a
$b$ quark and then decays to a $b\bar{b}$ pair, the octet pair production leads to a peak at a larger invariant 
mass of the leading and third jets.

Even more dramatic deformations of the distributions occur when the pair of $G_H$ scalars is produced through an $s$-channel 
resonance. 
We study these effects within a  renormalizable theory that includes an extension of the QCD gauge group, 
$SU(3)\times SU(3)$, as proposed in  \cite{Hill:1991at,Chivukula:1996yr}. 
The scalar sector responsible for spontaneously breaking this symmetry down to $SU(3)_c$ includes 
a $G_H$ scalar as well as two gauge singlet real scalar fields. We refer to this as the Renormalizable Coloron Model (ReCoM).
The heavy gauge boson arising from the extended symmetry, labelled by $G_\mu^\prime$ and referred to as the coloron (or more generally gluon-prime), 
is a color-octet particle of spin 1.  

We show that the coloron couples to a pair of $G_H$ scalars so that 
$s$-channel production of $G_\mu^\prime$ leads to a $4b$ signal. This is an example of nested resonances \cite{nested}: two pairs of 
$b$ jets form each a resonance and the combination of the two pairs also forms a resonance.
These multiple resonant features allow an efficient rejection of the background. However, for that it is necessary to have four $b$-tagged jets,
which requires large data sets. If only three $b$-tagged jets are required to pass the cuts, then the signal is larger and a detailed analysis could
separate it from the background within smaller data sets.

The CDF Collaboration has searched for a resonance in the invariant mass distribution of the two leading $b$ jets, produced in 
association with a third $b$ jet. Preliminary results \cite{HiggsPlusb} suggest the existence of a resonance 
with mass of about 140 GeV, that could be attributed to a fluctuation of the standard model 
background at the 6\% confidence level. A more puzzling feature of 
these CDF results  is that the measured transverse energy ($E_T$) distributions of the $b$ jets have shapes that differ notably from the standard 
model predictions \cite{CDF-bbb-url}. The shapes of the $E_T$ distributions of the two leading $b$ jets 
may be due to a fluctuation of the standard model only at around 1\% confidence level. We demonstrate that 
these shapes are nicely explained within the ReCoM when
the $G_H$ mass is 140 GeV and the $G_\mu^\prime$ mass is slightly above the $G_H G_H$ threshold.

We describe the interactions and decays of the spin-0 octet in Section~\ref{sec:color-octet}.
Then we analyze the Tevatron phenomenology of QCD-produced $G_H$ pairs in Section \ref{sec:Tevatron}. 
There we also discuss the boundstate  effects due to  gluon exchange between the two $G_H$ produced, and 
we compare the kinematic distributions of the multi-$b$-jet final state due to the scalar octet with those due to 
supersymmetric Higgs bosons. 
In Section \ref{sec:ReCoM} we derive the ReCoM predictions, compare them with the CDF data for several $3b$ kinematic distributions,
and discuss implications for other final states, such as those arising from associated production of a coloron and a $W$ boson.
We summarize strategies to distinguish different models with multi $b$ jets in Section~\ref{sec:conclusion}.

\section{Spin-0, weak-singlet, color-octet particle}\label{sec:color-octet}\setcounter{equation}{0}

The theory considered in this section is the standard model plus only one 
particle, $G_H$, which is a real field of spin 0, transforming as
an octet under the QCD gauge group $SU(3)_c$ and as a singlet under the 
electroweak gauge group $SU(2)_W \times U(1)_Y$. These gauge charges imply that 
$G_H$ is electrically neutral, and does not have any renormalizable interactions
with the quarks and leptons (note that $G_H \bar{b}_L b_R$ is not invariant under $SU(2)_W$; the opposite is true 
in the case of weak-doublet scalar octets \cite{Manohar:2006ga, Gresham:2007ri}). 
We refer to $G_H$ as a scalar octet, independently of whether it is a composite (as in technicolor)
or elementary particle.

\subsection{Renormalizable interactions}

The renormalizable couplings of $G_H$ to gluons are fixed by  $SU(3)_c$ gauge invariance:
\begin{eqnarray}
\frac{g_s^2}{2}f^{abc}f^{ade}\,G_\mu^b \, G^{\mu\,d}\, G_H^c\, G_H^e \,+ \, g_s\,f^{abc} \,G_\mu^a\, G_H^b\, \partial^\mu G^c_H  ~,  
\label{gluons}
\end{eqnarray}
where $f^{abc}$ is the anti-symmetric  $SU(3)_c$ tensor, $g_s$  is the QCD coupling, and $G_\mu$ is the gluon field.
The only other renormalizable couplings of $G_H$ are to the Higgs doublet ($H$) and to itself:
\be
\frac{\lambda_{HG}}{2} \, G_H^a G_H^a H^\dagger H + \frac{\lambda_G}{8}  \, (G_H^a G_H^a)^2 
+ \mu_G \, d_{abc} \, G_H^a G_H^b G_H^c ~~,
\label{quartic}
\ee
where $\lambda_{HG}$ and $\lambda_G>0$ are dimensionless parameters, $\mu_G$ is a parameter of mass dimension one,
and $d_{abc}$ is the totally-symmetric $SU(3)_c$ tensor.

The first term in Eq.~(\ref{quartic}) contributes to the mass squared of $G_H$ after electroweak symmetry breaking. We take the sum ($M_{G_H}^2$)
of this contribution and the mass squared from the Lagrangian to be positive. As a result,
$G_H$ does not have a vacuum expectation value (VEV) provided there is an upper limit on the cubic coupling, 
$|\mu_G| \lae \sqrt{\lambda_G}\,M_{G_H}$. We are primarily interested in the case where the physical mass of $G_H$, $M_{G_H}$, is above around 100 GeV.

The production of $G_H$ at hadron colliders occurs mainly in pairs, due to the couplings (\ref{gluons}) to gluons, via the tree-level diagrams shown in Fig.~\ref{fig:GG}. Single $G_H$ production is possible at one-loop through a cubic interaction [the last term in Eq.~(\ref{quartic})], but it is suppressed enough to be neglected.

\begin{figure}[t!]
\vspace*{-0.1cm}
\begin{center} 
{
\unitlength=0.78 pt
\SetScale{0.78}
\SetWidth{0.8}      
\normalsize    
{} \allowbreak
\begin{picture}(100,100)(5,0)
\ArrowLine(-10,80)(10,50)
\ArrowLine(10,50)(-10,20)
\Gluon(10,50)(65,50){3}{7}
\DashLine(65,50)(85,80){4}
\DashLine(65,50)(85,20){4}
\Text(40,62)[c]{\small $g$}
\Text( -17,80)[c]{\small $q$}
\Text( -17,20)[c]{\small $\bar{q}$}
\Text(100,80)[c]{\small $G_H$}
\Text(100,20)[c]{\small $G_H$}
\end{picture}
\quad\quad\quad
%
\begin{picture}(100,100)(8,0)
\Gluon(15,80)(38,50){-3}{4}
\Gluon(15,20)(38,50){-3}{4}
\Gluon(38,50)(70,50){3}{4}
\DashLine(70,50)(95,80){4}
\DashLine(70,50)(95,20){4}
\Text(54,62)[c]{\small $g$}
\Text( 8,80)[c]{\small $g$}
\Text( 8,20)[c]{\small $g$}
\Text(110,80)[c]{\small $G_H$}
\Text(110,20)[c]{\small $G_H$}
\end{picture}
\quad\quad
%
\begin{picture}(100,100)(-3,0)
\Gluon(15,80)(50,65){3}{4}
\Gluon(15,20)(50,35){3}{4}
\DashLine(50,65)(50,35){4}
\DashLine(50,65)(85,80){4}
\DashLine(50,35)(85,20){4}
\Text(35,50)[c]{\small $G_H$}
\Text( 8,80)[c]{\small $g$}
\Text( 8,20)[c]{\small $g$}
\Text(100,80)[c]{\small $G_H$}
\Text(100,20)[c]{\small $G_H$}
\end{picture}
\quad\quad
%
\begin{picture}(100,100)(-10,0)
\Gluon(15,80)(50,50){3}{4}
\Gluon(15,20)(50,50){3}{4}
\DashLine(50,50)(85,80){4}
\DashLine(50,50)(85,20){4}
\Text( 8,80)[c]{\small $g$}
\Text( 8,20)[c]{\small $g$}
\Text(100,80)[c]{\small $G_H$}
\Text(100,20)[c]{\small $G_H$}
\end{picture}
}
\end{center}
\vspace*{-0.8cm}
\caption{$G_H G_H$ production in hadronic collisions 
($u$-channel $G_H$ exchange is not shown). 
Curly lines represent gluons, 
while dashed lines represent scalar octets.} 
\label{fig:GG}
\end{figure}
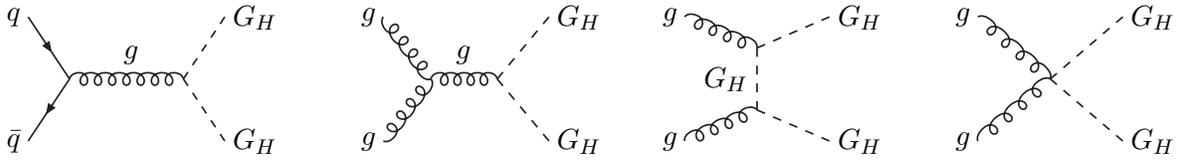

\begin{figure}[t!]
\vspace*{-0.1cm}
\begin{center} 
{
\unitlength=0.7 pt
\SetScale{0.7}
\SetWidth{0.8}      
\normalsize    
{} \allowbreak
\begin{picture}(100,100)(100,0)
\DashLine(15,30)(70,30){4}
\DashLine(70,30)(140,70){4}
\DashLine(70,30)(140,0){4}
\DashLine(140,0)(140,70){4}
\Gluon(140,0)(195,0){-3}{6}
\Gluon(140,70)(195,70){3}{6}
\Text(17,13)[c]{\small $G_H$}
\Text(100,67)[c]{\small $G_H$}\Text(100,0)[c]{\small $G_H$}
\Text(158,30)[c]{\small $G_H$}
\Text(208,0)[c]{\small $g$}\Text(208,70)[c]{\small $g$}
\end{picture}
\quad\quad\quad
%
\begin{picture}(100,100)(-8,0)
\DashLine(15,30)(70,30){4}
\DashCArc(105,30)(35,0,180){4}
\DashCArc(105,30)(35,180,0){4}
\Gluon(140,30)(195,60){3}{6}
\Gluon(140,30)(195,0){3}{6}
\Text(17,13)[c]{\small $G_H$}
\Text(70,67)[c]{\small $G_H$}\Text(70,-4)[c]{\small $G_H$}
\Text(207,0)[c]{\small $g$}\Text(207,60)[c]{\small $g$}
\end{picture}
}
\end{center}
\vspace*{0.1cm}
\caption{Scalar octet decay to gluons, due to the trilinear $G_H$ interaction of  Eq.~(\ref{quartic}).
A diagram similar with the left one but with interchanged end points for the gluon lines is not shown.
}
\label{fig:loops}
\end{figure}
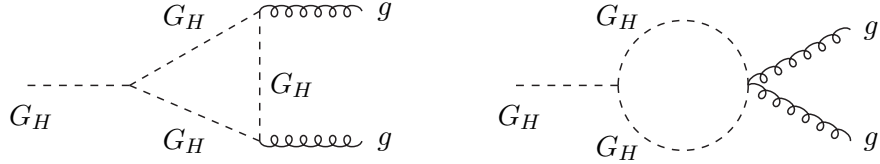

\subsection{Decays of the scalar octet}

The only renormalizable coupling of $G_H$ that violates the $Z_2$ invariance under the 
$G_H \to - G_H$ transformation is the cubic term in  Eq.~(\ref{quartic}).
Thus, the only decays of $G_H$ allowed by the renormalizable couplings shown above 
occur at one or more loops and involve at least one vertex proportional to $\mu_G$.
At one-loop, divergent terms from the triangle and bubble diagrams shown in Fig.~\ref{fig:loops}
cancel each other, and the finite result for the width of
$G_H$ decaying into two gluons is
\be
\Gamma_0 (G_H \to gg) = \frac{15 \,\alpha_s^2 \, \mu_G^2 }{128 \,\pi^3 \,M_{G_H}}\,\left(\frac{\pi^2}{9} -1 \right)^{\! 2}  \; ~,
\label{eq:decaygg}
\ee
where $\alpha_s$ is the QCD coupling evaluated at $M_{G_H}$. 
This width is accidentally suppressed by the small numerical coefficient given in the paranthesis 
(see \cite{Gresham:2007ri} for a similar case).

Decays of $G_H$ into quarks induced by nonrenormalizable couplings may have large branching fractions
because the above decay width into gluons is loop-suppressed. 
Dimension-5 operators of this type involve either a Higgs doublet,
\be
\frac{C^d_{ij}}{m_\psi} H G_H^a \overline{Q}_L^i T^a d_R^j + \frac{C^u_{ij}}{m_\psi} \widetilde{H} G_H^a \overline{Q}_L^i T^a u_R^j + {\rm H.c.}  ~~,  
\label{dim-5} 
\ee
or a covariant derivative,
\be
\frac{1}{m_\psi}
\left(  K^d_{ij}\, \overline{d}_R^i \gamma^\mu T^a d_R^j  + K^u_{ij}\, \overline{u}_R^i \gamma^\mu T^a u_R^j  + K^Q_{ij}\, \overline{Q}_L^i \gamma^\mu T^a Q_L^j \right) D_\mu G^a_H + {\rm H.c.}   ~~, 
\label{derivative-dim-5} 
\ee
where $Q_L^i \equiv (u_L^i, d_L^i)$, the indices $i,j =1,2,3$ label the generations, $C^{d, u}_{ij}$ and $K^{d, u, Q}_{ij}$ are complex coefficients, and 
$m_\psi$ is the mass of a heavy particle which generates these operators (if 
several particles of different masses contribute to these operators, then their effects can be included in the coefficients).
Note that the derivative parts of operators (\ref{derivative-dim-5}) may be transformed into the operators (\ref{dim-5}), 
with the Higgs field replaced by its VEV, through integration by parts and the use of quark field equations. The non-derivative parts of the 
three operators only mediate 3-body decays, which are less important compared to the 2-body decays. 

There are also dimension-5 operators contributing to the $G_H \to gg$ decay,
\be
\frac{\alpha_s}{4\pi m_\psi} d_{abc} \, G_H^a \, G^b_{\mu\nu} \left( \xi_G \, G^{c\,\mu\nu} + \xi_G^\prime \, \epsilon^{\mu\nu\alpha\beta}G^{c}_{\alpha\beta} \right) ~~,
\label{psi-gg}
\ee 
and to the $G_H \to gZ$ or $g\gamma$ decays,
\be
\frac{\sqrt{\alpha_s\,\alpha }}{4\pi \cos\theta_W \, m_\psi}\,G_H^a\,G^a_{\mu\nu}\left( \xi_B B^{\mu\nu}  +  \xi_B^\prime \epsilon^{\mu\nu\alpha\beta} B_{\alpha\beta}\right)  ~~.
\label{psi-gZ}
\ee
Here, $B_{\mu\nu}$ is the field tensor of the $U(1)_Y$ gauge boson ($-\sin\theta_W Z^\mu + \cos\theta_W A^\mu$, where $A^\mu$ is the photon field) 
and $\theta_W$ is the weak mixing angle. These operators are generated by loops involving some new particle of mass $m_\psi$ whose coupling to $G_H$ determines the dimensionless coefficients $\xi_G$, $\xi_G^\prime$,
$\xi_B$ and $\xi_B^\prime$.
The decay of a scalar octet into a gluon and a photon has been studied in 
Ref.~\cite{Bai:2010mn}.

Operators that allow $G_H$ to decay into a Higgs boson and gluons arise at dimension-7 or higher. For example, the operator $(D^\mu G_H^a)G^a_{\mu\nu}\,H^\dagger D^\nu H$ induces the $G_H \to g\,g\,h$ decay (the $G_H \to g\,h$ process is forbidden by angular momentum conservation). The operators that 
couple $G_H$ to leptons also appear only at dimension-7 or higher, for example, $(D^\mu G_H^a)\,\overline{L^i}\gamma^\nu L^j\,G^a_{\mu\nu}$. 

\begin{figure}[t!]
\vspace*{-0.1cm}
\begin{center} 
{
\unitlength=1 pt
\SetScale{1}
\SetWidth{1}      
\normalsize    
{} \allowbreak
\begin{picture}(100,80)(60,-30)
\ArrowLine(60,-20)(0,0)
\ArrowLine(0,0)(40,20)\ArrowLine(40,20)(80,40)\ArrowLine(80,40)(140,50)
\DashLine(-50,0)(0,0){4}
\DashArrowLine(10,50)(80,40){3}
\Line(41,25)(39,15)\Line(45,19)(35,21)
\Text(-40,-13)[c]{\small $G_H$}\Text(60,-30)[c]{\small $b_R$}\Text(145,40)[c]{\small $b_L$}
\Text(40,5)[c]{\small $\psi_{L}$}\Text(70,20)[c]{\small $\psi_{R}$}
\Text(20,60)[c]{\small $\langle H \rangle$}
\quad\quad\quad
\end{picture}
\begin{picture}(100,100)(15,0)
\DashLine(30,30)(75,30){4}
\ArrowLine(140,70)(75,30)
\ArrowLine(75,30)(140,0)
\ArrowLine(140,0)(140,70)
\Gluon(140,0)(195,0){-3}{6}
\Gluon(140,70)(195,70){3}{6}
\Text(38,18)[c]{\small $G_H$}
\Text(100,60)[c]{\small $\psi$}\Text(100,4)[c]{\small $\psi$}
\Text(153,30)[c]{\small $\psi$}
\Text(208,0)[c]{\small $g$}\Text(218,70)[c]{\small $g, \gamma, Z$}
\end{picture}
\
}
\end{center}
\caption{Effective couplings of a scalar octet to $b$ quarks and the Higgs VEV (left diagram),
or to a gluon pair (right diagram),
induced in the presence of a heavy vectorlike quark $\psi$ and leading to the $G_H \to b\bar{b}, gg, g\gamma, gZ$ decays.} 
\label{fig:B}
\end{figure}
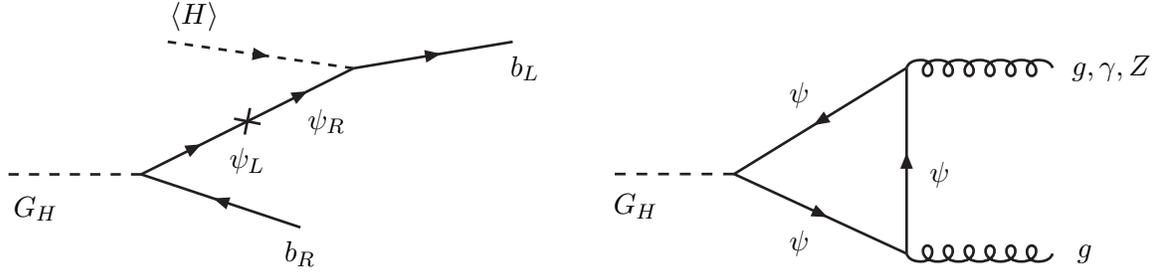

Given that the operators (\ref{dim-5}) involve the Higgs doublet, their coefficients are typically proportional to 
the quark mass. It is natural to assume that the dominant decay mode induced by 
operators (\ref{dim-5})
is $G_H \to t\bar{t}$ for $M_{G_H}\!\! \gae 350$ GeV, 
and $G_H \to b\bar{b}$  for 10 GeV$\lae M_{G_H}\!\! \lae 350$ GeV. However, the relative normalization of the 
$C^d_{ij}$ and $C^u_{ij}$ coefficients depends on the underlying mechanism responsible for generating these operators.

To be concrete, we consider the following renormalizable model as an origin for the dimension-5 operators.
A vectorlike quark, $\psi$, having the same gauge charges as $b_R$ may mix with the down-type 
quarks via the following terms in the Lagrangian:
\be
m_\psi \overline{\psi}_{L} \psi_{R} + \mu_{i\psi} \overline{\psi}_{L} d_R^i + 
y_{ij} H \overline{Q}_L^i d_R^j + \lambda_{i} H \overline{Q}_L^i \psi_{R} + {\rm H.c.}~,
\label{Yukawa}
\ee
where $y_{ij}$ and  $\lambda_{i}$ are Yukawa couplings, $\mu_{i\psi}$ is a mass mixing parameter, 
and $m_\psi$ is the $\psi$ fermion mass in the small $\mu_{i\psi}$ limit.
The scalar octet also has Yukawa couplings to $\psi$:
\be
G_H^a \left( \eta_0 \, \overline{\psi}_{L} T^a \psi_{R} + \eta_i \, \overline{\psi}_{L}  T^a  b_R^i \right) + {\rm H.c.} 
\label{GH-Yukawa}
\ee
The above interactions lead to effective couplings of $G_H$ to standard model quarks,
as shown in the first diagram of Fig.~\ref{fig:B}.  
Upon integrating $\psi$ out, the operators (\ref{dim-5}) are induced with coefficients
\be
C^u_{ij} = 0  \;\;\;\; , \;\;\;\; C^d_{ij} \approx \lambda_i\, \eta^*_j ~~,
\ee
where we assumed for simplicity that $m_\psi \gg \mu_{i\psi}$.
The coefficients $C^d_{ij}$ are defined in the gauge eigenstate basis. 
Assuming that the two unitary matrices that diagonalize the down-type quark mass matrix are close to the identity matrix, the $G_H$ couplings to down-type quarks in the mass eigenstate basis are approximately equal to $C_{ij}^d$. For $|\lambda_3| \gg |\lambda_1|,|\lambda_2|$ and $|\eta_3| \gg |\eta_1|,|\eta_2|$ the width for $G_H \to b\bar{b}$ is much larger than for any other fermion final states, and is given by
\be
\Gamma \left(G_H \to b\bar{b} \right) \simeq  \frac{|\lambda_3 \eta_3|^2 \, v^2}{24\pi \, m_\psi^2 } M_{G_H} ~~,
\label{eq:decaybb}
\ee
where $v = 246$ GeV is the electroweak scale.

The $\psi$ fermion also induces some of the dimension-5 operators given in Eqs.~(\ref{psi-gg}) and (\ref{psi-gZ}),
as shown in Fig.~\ref{fig:B}. The ensuing decay width into gluons, which adds to $\Gamma_0$ given 
in Eq.~(\ref{eq:decaygg}), is 
\be
\Gamma_\psi \!\left(G_H \to g g \right) \simeq \frac{5 \, \eta_0^2 \, \alpha_s^2 \, M_{G_H}^3}{6 \, (16\pi)^3 \, m_\psi^2} \left|A\left(M_{G_H}^2/(4m_\psi^2)\right)\right|^2 ~~,
\ee
where the function $A$ arises from a momentum integral given in \cite{Rizzo:1979mf}:
\be
A(x) = \frac{2}{x^2} \left[ x + (x - 1) {\rm Arcsin}^2 \sqrt{x}\;  \right] \;\; , \;\;
{\rm for} \; \; x \le 1 ~~.
\ee
For $x \ll 1 $, {\it i.e.} $m_\psi \gg M_{G_H}/2$, the function has the value 
$A \approx 4/3$, and the coefficients of the operators (\ref{psi-gg}) are 
\be
\xi_G = \frac{\eta_0}{6\sqrt{2}}   \;\;\;\; , \;\;\;\; \xi_G^\prime = 0 ~~.
\label{eq:coef}
\ee

The ratio of the decay widths into $b\bar{b}$ and $gg$, neglecting the interference of the two diagrams contributing to $G_H \rightarrow gg$ 
in Figures~\ref{fig:loops} and \ref{fig:B}, is given by 
\be
\frac{\Gamma \left(G_H\! \to b\bar{b}\right)}
{\Gamma_\psi \!\left(G_H\! \to gg\right) + \Gamma_0 \left(G_H\! \to gg\right)}
= \frac{0.69}{\eta_0^2 + 3.02 \, \mu_G^2 m_\psi^2 / M_{G_H}^4 } \left(\frac{|\lambda_3 \eta_3|}{10^{-3}}
\, \frac{100 \; {\rm GeV}}{M_{G_H}}\, \frac{0.1}{\alpha_s} \right)^{\! 2} ~~.
\ee
 Note that
$m_\psi$ may be significantly above the TeV scale, so that the vectorlike quark might not be produced even at the LHC. Thus, the branching fraction of $G_H \to b\bar{b}$ can  be very small if $\mu_G m_\psi \gg M_{G_H}$.
Alternatively, if $\psi$ is not much heavier than $G_H$, and $\eta_0 = O(1)$,
then the branching fraction of $G_H \to b\bar{b}$ can be large only for
$|\lambda_3 \eta_3| \gae 10^{-3}$.

The scalar octet has a very narrow width (several orders of magnitude less than its mass), 
as its contributions arise from loops or higher dimensional operators.
Its decays are usually prompt, unless $\mu_G$ is very small and $m_\psi$ is very large. For example,
$\mu_G = 0$, $\eta_0 = 0$ and $m_\psi/|\lambda_3 \eta_3| \gae 2.5\times 10^5$~TeV give a decay length 
$c\tau \gae 100~\mu m$ so that $G_H$ decays off the beam line. We will not investigate further these 
signatures involving dijet and $b\bar{b}$ resonances originating from displaced vertices.

The decays $G_H \to g \gamma, g Z$ have small branching fractions due to coupling and color suppressions
[the coefficients of the operators (\ref{psi-gZ}) are comparable with those in Eq.~(\ref{eq:coef}):
$\xi_B = \xi_G/3$ and $\xi_B^\prime = 0$].
For example, 
\be
\frac{\Gamma \left(G_H \to g \gamma\right)}{\Gamma_\psi \!\left(G_H \to g g \right)} 
= \frac{\alpha}{15 \alpha_s} \approx 0.44\%  ~~.
\ee
Nevertheless, a photon-plus-jet resonance that arises together with a dijet resonance
of equal mass, is an interesting 
signature that follows from the decays of a $G_H$ pair.

\subsection{CP violation in $B_s$ mixing from scalar octet exchange}

The operators (\ref{dim-5}) can induce tree-level flavor-changing neutral-current (FCNC) processes mediated by the scalar octet. In particular, 
integrating out the octet gives the following $\Delta B = 2$ terms in the effective Lagrangian:
\be
\frac{v^2}{2\,M_{G_H}^2 m_\psi^2}\, \left( \eta_3^\prime \lambda_2^\prime \, \overline{b}_L T^a s_R + \eta_2^\prime \lambda_3^\prime \, \overline{b}_R T^a s_L\right)^2 ~~,
\label{bs}
\ee
where the quark fields shown here are mass eigenstates, and $\eta_i^\prime$ and $\lambda_i^\prime$ are the Yukawa couplings of Eqs.~(\ref{Yukawa}) and 
(\ref{GH-Yukawa}) transformed to the mass eigenstate basis.

Tree-level exchange of octet bosons has been proposed before   \cite{Dobrescu:2010rh} as a possible explanation for
the anomalous dimuon charge asymmetry reported by the D0 Collaboration~\cite{Abazov:2010hv}. The novel feature in our case is that 
the scalar octet  has dimension-5 couplings to the standard model quarks. Nevertheless, we now show that the effect of $G_H$ is large enough
if the $G_H$ mass is below a few hundred GeV.

Using the relation $T^a_{ij}T^a_{kl} = - \delta_{i}^{j}\delta_{k}^{l}/6 + \delta_{i}^{l}\delta_{j}^{k}/2$, and substituting the matrix elements of the ensuing operators from Ref.~\cite{Becirevic:2001xt}, we
find the matrix element of the Hamiltonian associated with operator (\ref{bs}): 
\bear
\langle \overline{B}_s | {\cal H}^{\rm NP} | B_s \rangle  = 
\frac{v^2 M^4_{B_s}f^2_{B_s}}{96\,M_{G_H}^2m_\psi^2(m_b + m_s)^2} && \hspace*{-1.2em} \left[  \left(  \rule{0mm}{4mm} (\eta_2^\prime \lambda_3^\prime)^2 
+ (\eta_3^\prime \lambda_2^\prime)^2 \right) \left(\!B_3 + \frac{5}{3}B_2\!\right) \right.
\nonumber \\
 && \left. + \, 4
\eta_3^\prime \lambda_2^\prime \eta_2^\prime \lambda_3^\prime (B_5 - B_4) \rule{0mm}{4.5mm} \right] ~~.
\label{eq:M12NP}
\eear
The decay constant, computed on the lattice with 2+1 flavors \cite{Gamiz:2009ku}, is $f_{B_s} = 231\pm 15$~MeV.
The ``bag" parameters have been estimated on the lattice in the quenched approximation \cite{Becirevic:2001xt}:
$B_2 = 0.80$, $B_3 = 0.93$, $B_4 = 1.16$ and $B_5=1.75$. 

If  $\eta_3^\prime \lambda_2^\prime$ or $\eta_2^\prime \lambda_3^\prime$ have complex phases, then 
the $G_H$ exchange induces CP violation in $B_s-\bar{B}_s$ mixing.
Parametrizing
\bear
\langle \overline{B}_s | {\cal H}^{\rm NP} | B_s \rangle  \equiv \left(C_{B_s} e^{-i\phi_s} - 1\right) 2M_{B_s} (M_{12}^{\rm SM})^*
\eear
where $M_{12}^{\rm SM} \simeq (9.0 \pm 1.4)$ ps$^{-1}$ \cite{Lenz:2006hd, Dobrescu:2010rh}, and using Eq.~(\ref{eq:M12NP}) 
we find that the vectorlike quark mass required to induce the CP-violating phase $\phi_s$ is 
\be
m_\psi = \frac{v \,  f_{B_s} M_{B_s}^{3/2}  \left|2.3 \left[ (\eta_2^\prime \lambda_3^\prime)^2 
+ (\eta_3^\prime \lambda_2^\prime)^2 \right] + 2.4\,\eta_3^\prime \lambda_2^\prime \eta_2^\prime \lambda_3^\prime \rule{0mm}{4mm} \right|^{1/2} } 
{ 8 M_{G_H} (m_b + m_s) \left(3 M_{12}^{\rm SM} \right)^{1/2}\left(C_{B_s}^2 + 1 - 2 C_{B_s} \cos{\phi_s}  \right)^{\! 1/4} } ~~.
\ee
The measured $B_s$ mass difference gives the constraint $C_{B_s} = 0.98 \pm 0.15$, and the D0 dimuon asymmetry requires 
$\phi_s$ near $- \pi/2$. For illustration, let us assume $\eta_3^\prime \lambda_2^\prime = \eta_2^\prime \lambda_3^\prime$,
$C_{B_s} = 1$ and $\phi_s = -\pi/2$, which gives
\be
m_\psi\,\approx\, 1.1 ~\mbox{TeV} \; \frac{|\eta_3^\prime \lambda_2^\prime|}{10^{-2}}\,\frac{100~\mbox{GeV}}{M_{G_H}} 
~~.
\ee
Thus, for $|\eta_3^\prime \lambda_2^\prime| = |\eta_2^\prime \lambda_3^\prime| \sim 10^{-2}$, the D0 dimuon asymmetry suggests that the 
vectorlike quark has a mass within the reach of the LHC.

\section{Tevatron phenomenology of scalar octets} 
\label{sec:Tevatron}\setcounter{equation}{0} 

\begin{figure}[h!]\center
\hspace*{-0.9em}
\psfig{file=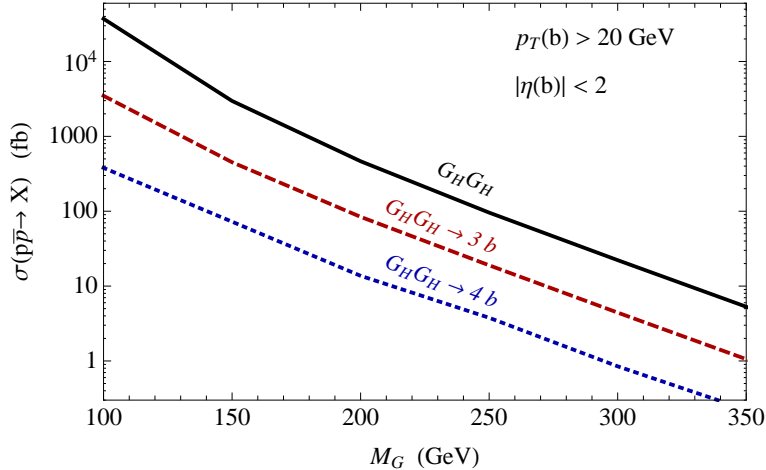,width=10.1cm,angle=0}
\vspace*{-2mm}
\caption{Leading-order cross section for production of a  $G_H$ pair (solid line) at the Tevatron. Also shown are the cross sections after cuts 
for producing a pair of $G_H$ followed by each decaying with 100\% branching fraction into $b\bar{b}$ and requiring that 
3 (dashed line) or 4 (dotted line) jets pass the cuts and are $b$-tagged. }
\label{fig:cross-section}
\end{figure}

Scalar octets are produced in pairs at hadron colliders (see Fig.~\ref{fig:GG})
and decay with a large branching fraction into $b\bar{b}$. The partonic cross sections are given by~\cite{Chivukula:1991zk, Manohar:2006ga}.
\bea
\sigma(q \bar{q} \rightarrow G_H G_H)&=& \frac{2\,\pi\,\alpha_s^2}{9\,\hat{s}}\, \beta^3\,, \\
\sigma(g g \rightarrow G_H G_H)&=& \frac{3\,\pi\,\alpha_s^2}{32\,\hat{s}}\,\left[ 27\,\beta - 17\,\beta^3 + 3\,(\beta^4 + 2\beta^2 -3)\ln\left( \frac{1+\beta}{1-\beta} \right) \right]\,,
\eea
where $\beta = (1- 4 M_{G_H}^2/\hat{s})^{1/2}$ is the velocity of the octet in the center of mass frame. 
Thus, the signal consists of a pair of narrow $b\bar{b}$ resonances of same mass, $M_{G_H}$. 
Including the interactions (\ref{quartic}) into the MadGraph/MadEvent
\cite{Alwall:2007st} package, 
we find that the processes shown in Fig.~\ref{fig:GG}, convoluted with the CTEQ 6L1
\cite{Pumplin:2002vw} parton distribution functions (PDFs), give the cross section for 
$p\bar{p} \to G_HG_H$ at the Tevatron shown by the solid line in Fig.~\ref{fig:cross-section}. 
The factorization and renormalization scales used by default in MadGraph/MadEvent
are equal and depend on the event: $(M_{G_H}^2\! + p_T^2(G_{1}))^{1/2}(M_{G_H}^2\! + p_T^2(G_{2}))^{1/2}$, 
where $p_T(G_{1})$ and $p_T(G_{2})$ are the transverse momenta of the two $G_H$ produced in each event.
This cross section depends only on $M_{G_H}$,
and is consistent with the result shown in Fig.~3 of Ref.~\cite{Dobrescu:2007yp}.

\subsection{3$b$ versus 4$b$ signals}

The QCD background is dramatically reduced by selecting
events containg four jets, all $b$ tagged, and then imposing that two of them have an
invariant mass close to that of the other two $b$ jets \cite{Dobrescu:2007yp}.
We point out, however, that the sensitivity to the pair of 
octets could be improved if, instead of requiring four $b$ jets, only the three
jets of largest transverse momenta are $b$ tagged and required to pass the basic cuts. In this case only two of the $b$ jets will form a resonance,
and therefore the background is substantially higher than in the case of a pair of resonances
of same mass. However, the signal is also larger, such that the ratio $S/\sqrt{B}$ where
$S$ and $B$ are the number of signal and background events, respectively, may be increased.

To see this, first note that the $b$ tagging efficiency is about 50\%, leading to a factor of 2 suppression 
of the $4b$ signal compared to the $3b$ signal. The $4b$ signal is further suppressed because
the 4th jet is rather soft, especially for low $M_{G_H}$, and does not always pass the basic cuts, such as $p_T > 20$ GeV, 
imposed by the CDF \cite{HiggsPlusb} and D0  \cite{Abazov:2010ci}  searches. To estimate the ensuing suppression 
of the $4b$ signal, we use MadGraph/MadEvents to generate 
parton-level events at tree level for $p\bar{p} \to G_H G_H \to b\bar{b}b\bar{b}$ at a center-of-mass energy of 1.96 TeV. 
Imposing the basic cuts used by CDF \cite{HiggsPlusb}, namely $p_T > 20$ GeV and $|\eta| < 2$ for each $b$ jet, 
we obtain the result shown in columns 2 and 3 of Table 1.

\begin{table}[b!]\small
\vspace*{4mm}
\renewcommand{\arraystretch}{1.2}
\centerline{
\begin{tabular}{|c||cc|cc|cc|}
\hline
& \multicolumn{2}{c|}{partonic level} &   \multicolumn{2}{c|}{radiation included} &  \multicolumn{2}{c|}{$b$-tagging included}  \\ \cline{2-7}
$M_{G_H}$ (GeV)     & 3$j$ & $4j$  &   \ \ $3j$ & $4j$  &    \ \ $3b$ & $4b$      \\ \hline
100 &  \ 91.1 & 62.1 &     \ \ \ 76.9 & 38.9 &   \ \ \   9.4 & 1.0   \\
150 &  \ 95.8 & 78.1 &     \ \ \ 93.2 & 65.5 &    \ \ \ 15.2 & 2.4   \\
200 &  \ 97.4 & 83.1 &     \ \ \ 96.9 & 77.8 &    \ \ \ 18.1 & 3.0   \\
250 &  \ 98.2 & 85.0 &     \ \ \ 98.4 & 84.3 &    \ \ \ 19.8 & 3.9   \\
300 &  \ 98.6 & 86.8 &     \ \ \ 99.0 & 87.8 &    \ \ \ 19.9 & 3.8   \\
350 &  \ 98.9 & 86.7 &     \ \ \ 99.4 & 90.6 &    \ \ \ 20.0 & 4.2   \\ \hline 
\end{tabular}
}
\caption{Efficiency in \% for the three and four jets to pass the basic cuts ($p_T > 20$ GeV, $|\eta| < 2$): at parton level (columns 2,3), 
after radiation and detector effects included through {\sc Pythia} and PGS (columns 4,5),
and after $b$ tagging by PGS (columns 6,7).}
\label{tab:charges}
\end{table}

Initial and final state radiation further soften the 4th jet. We estimate this effect by processing
the parton-level events with {\sc Pythia} \cite{pythia} for showering and hadronization, and PGS~\cite{pgs} (with the default CDF detector card) 
for detector simulation. The result is shown in columns 4 and 5 of Table 1.
The $b$-tagging efficiency also decreases for softer jets, so that the efficiency for four $b$ jets to pass the basic cuts 
is significantly smaller than that for three $b$ jets, especially for low $M_{G_H}$ (see the last column of Table 1).
In Fig.~\ref{fig:cross-section} we show the signal cross sections for 3 and 4 $b$-tagged jets after basic kinematic cuts.

\subsection{Properties of the $3b$ signal}

Given that the 3$b$ signal is larger than the $4b$ signal by a factor ranging between 9 and 5 when $M_{G_H}$ varies between 100 and 350 GeV
(see last two columns of Table~\ref{tab:charges}), 
it is useful to focus on the  3$b$ final state and analyze its special features, which may be used for reducing the background.
The most clear feature of this signal is that two of the three $b$ jets form a narrow resonance of mass close to $M_{G_H}$. 
Let us label the three $b$ jets by $b_i$, $i = 1,2,3$. Within each event we take $b_1$, $b_2$, $b_3$ to be the jets of largest, 
2nd-largest and 3rd-largest transverse momenta ($p_T$), respectively.
In the limit where $M_{G_H}$ is very large, close to $\sqrt{s}/2$,
the two octets are produced mostly at rest and the $p_T$ of a jet is fixed by the 
angle between the two $b$ jets from the decay of an octet such that $b_1$ and $b_2$ form a mass peak.

\begin{figure}[t]\center
\hspace*{-0.9em}
\psfig{file=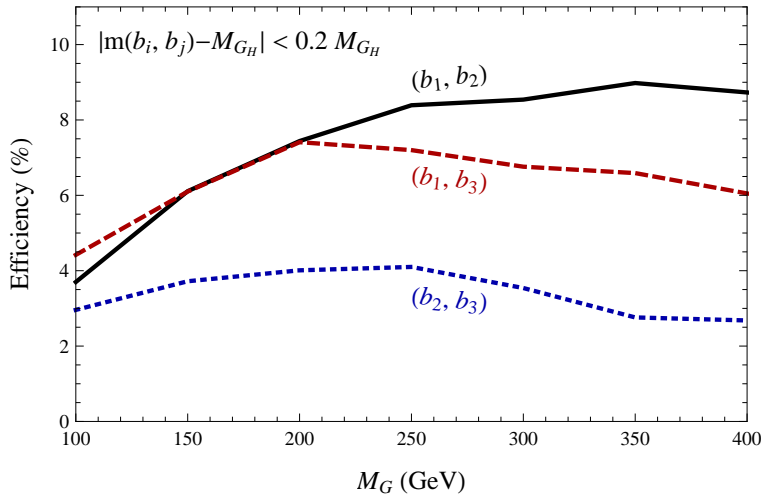,width=10.1cm,angle=0}
\vspace*{-1mm}
\caption{Fraction of signal events having the invariant mass of the $i$th and $j$th jets, $m_{ij} \equiv m(b_i, b_j)$, within 20\% of the octet mass $M_{G_H}$.
The fractions for combinations with the 4th jet are below $1\%$ and are not shown here.}
\label{fig:fractions}
\end{figure}

For masses of interest at the Tevatron, between 100 and 400 GeV, the $p_T$'s of the octets are comparable with the 
$p_T$'s of the jets in the rest frame of the corresponding octet.
As a result, the mass peak is formed in some events by $b_1$ and  $b_2$ while in other events by 
$b_1$ and  $b_3$, or even by the $b_2$ and $b_3$ jets. We refer to these three types of events as $(b_1,b_2)$,  $(b_1,b_3)$, and  $(b_2,b_3)$
respectively. To estimate the ratios between the numbers of events falling into these three categories, we impose an invariant 
mass cut $|m_{ij} - M_{G_H} | < 0.2 M_{G_H}$, where $m_{ij} \equiv m(b_i, b_j)$ is the invariant mass of the $b_i$ and $b_j$ jets.
We show the efficiency of the combination of basic and invariant mass cuts in Fig.~\ref{fig:fractions}.
Given that the $(b_1, b_2)$ and $(b_1,b_3)$ events have comparable efficiencies, it is  useful to use both $m_{12}$ and $m_{13}$ as 
discriminants between the octet signal and the QCD background.
Note that the ratio of the efficiencies for $(b_1, b_2)$ and $(b_1,b_3)$ events increases with $M_{G_H}$,
which is explained by the large $M_{G_H}$ argument presented above.

So far there have been no collider searches for the pair-produced octet. However, the CDF and D0 Collaborations have 
searched for a 3$b$ signal predicted in the MSSM at large $\tan\beta$: a $b$ quark from the proton (in the 5-flavor PDF scheme) radiates off a heavy Higgs
boson which then decays into a $b\bar{b}$ pair. This is the same final state as the 3$b$ one due to octets discussed here, but we will show that the kinematic distributions are different.

The CDF search \cite{HiggsPlusb}, with 2.2 fb$^{-1}$ of data, shows an intriguing excess of events with $m_{12}$ in the 125--155 GeV range.
The probablity for this excess to arise from a fluctuation of the standard model background is 0.9\%,
and increases to 5.7\% when the whole range of invariant masses is taken into account.

The D0 search \cite{Abazov:2010ci}, with 5.2 fb$^{-1}$ of data, rules out the presence of MSSM Higgs bosons with couplings large enough to account 
for the CDF excess. The tension between the CDF and D0 results may be due to a statistical fluctuation. An alternative explanation, however, 
is that the D0 search is less sensitive to the octet-induced signal; this is a consequence of the optimization of the signal within the D0 search 
for the MSSM Higgs bosons through the use of a likelihood discriminant.
Some of the kinematic variables included in the likelihood discriminant ({\it e.g.}, the angular separation $\Delta\phi$ between the 
two $b$ jets that are most likely to originate from the decay of the new particle) may discriminate against events due to $G_H G_H$ 
production.

\begin{figure}[t]
\begin{center}
\includegraphics[width=0.48\textwidth]{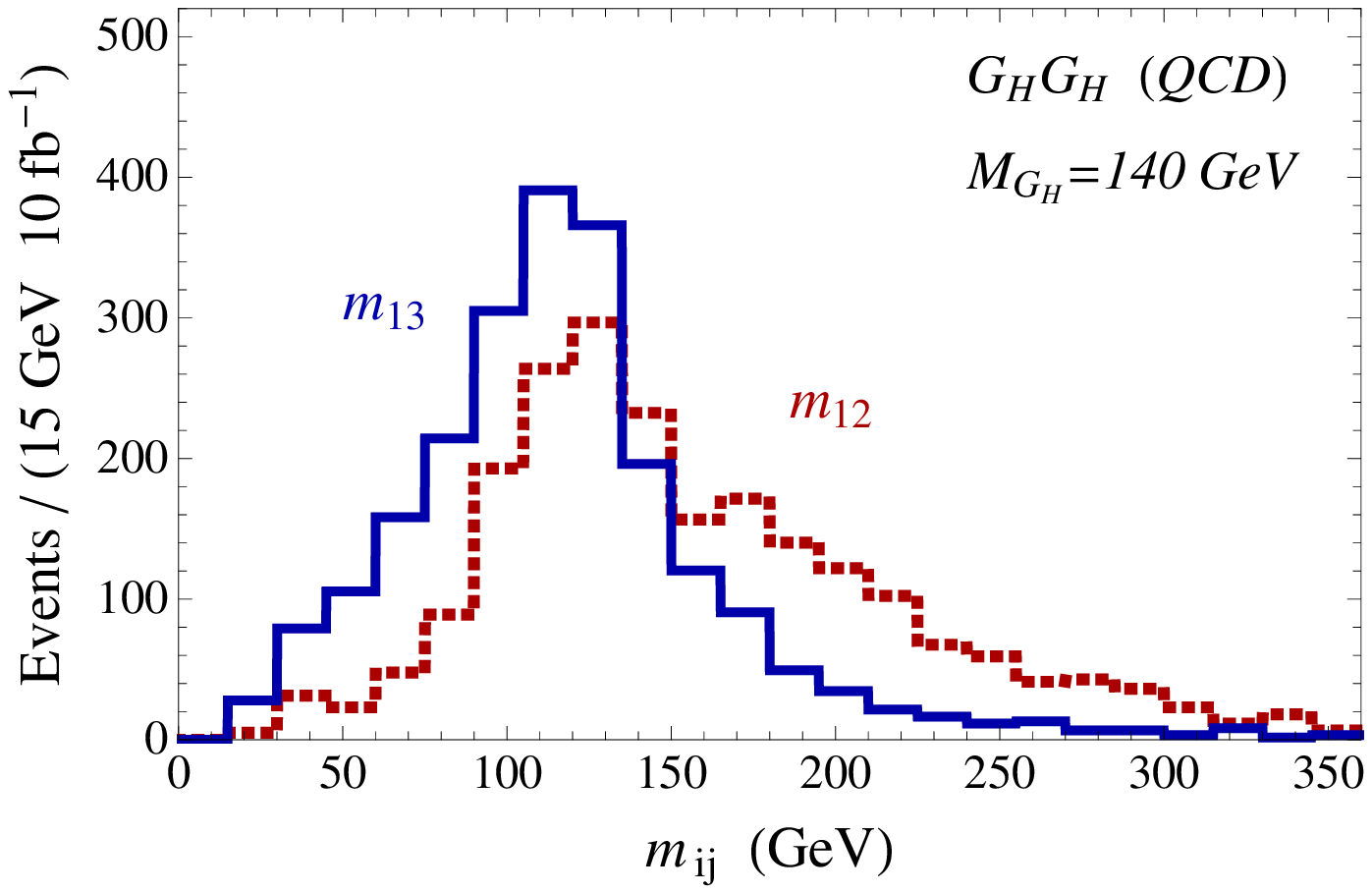} \hspace{2mm}
\includegraphics[width=0.48\textwidth]{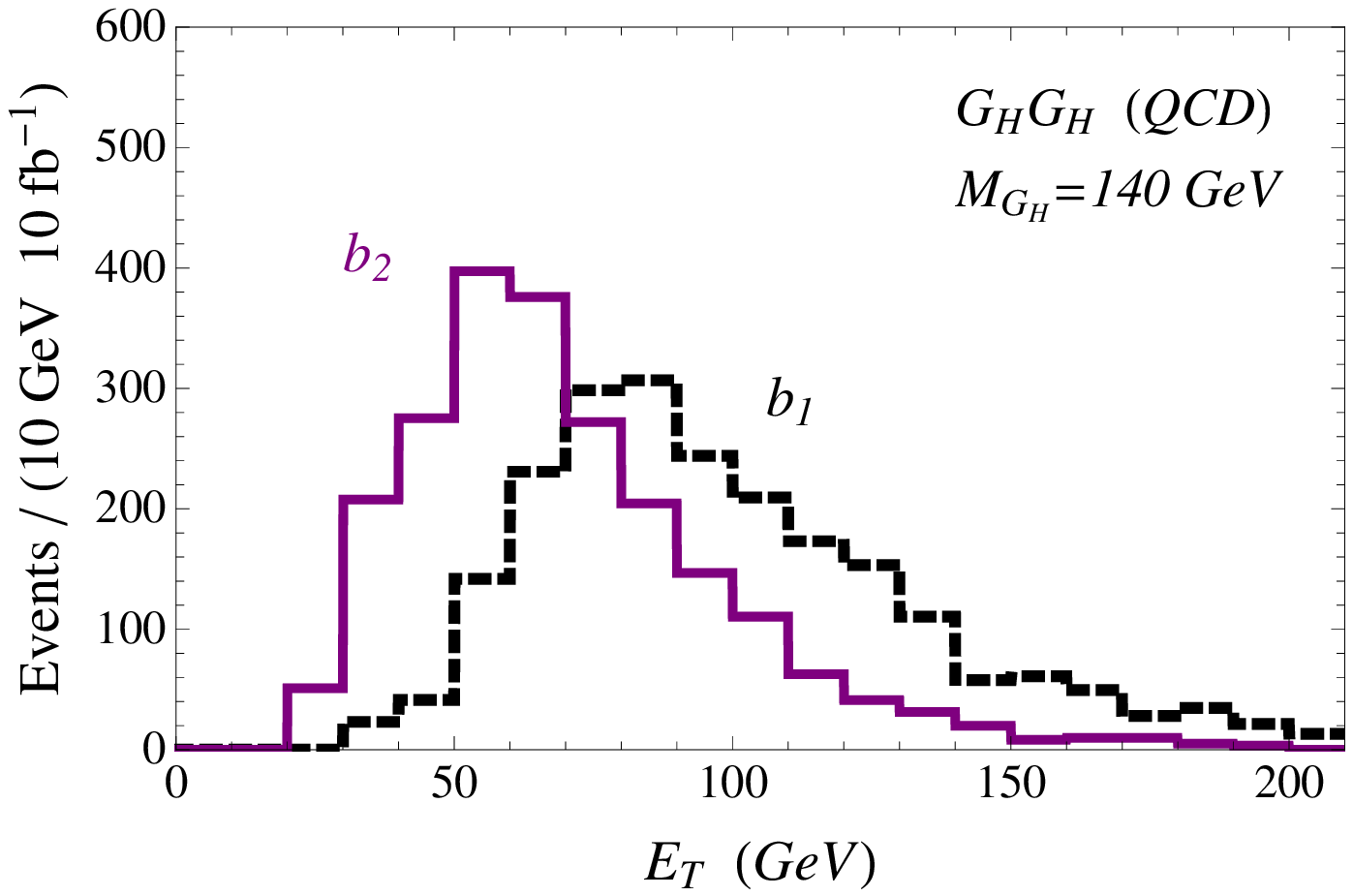}
\caption{Kinematic distributions resulting from $G_HG_H$  production at the Tevatron due to gluon couplings, for an octet mass of 140~GeV (cross 
section is 4.6~pb) and a $G_H \!\to b\bar{b}$ branching fraction of  60\%.
Left panel:  invariant mass distributions of the two leading $b$ jets (dotted red line), and of the 1st and 3rd leading $b$ jets (solid blue line).
Right panel: transverse energy distributions of the 1st (dashed black line) and 2nd (solid purple line) leading $b$ jets. }
\label{fig:mijEt}
\end{center}
\end{figure}

Motivated by the CDF search,  we first study the kinematics of an octet having a mass $M_{G_H} = 140$ GeV, and 
impose the basic cuts $p_T > 20$ GeV and $|\eta| < 2$ for each $b$ jet.  In Fig.~\ref{fig:mijEt}, left panel,
we show the number of events as a function of invariant masses of $b_1$ and $b_2$ ($m_{12}$), and of $b_1$ and $b_3$ ($m_{13}$).
Although the peak of the $m_{12}$ distribution is close to  $M_{G_H}$, the spectrum is broad such that our mass 
window cut $|m(b_1, b_2) - M_{G_H}| < 0.2 M_{G_H}$ keeps only around 50\% of signal events.
The $m_{13}$ distribution also shows a clear peak, which is narrower than the $m_{12}$ one.
The $m_{13}$ peak is at a lower invariant mass than the $m_{12}$ peak, because radiation effects make the 3rd jet softer.
The transverse energy distributions of the two leading jets (see the right panel of Fig.~\ref{fig:mijEt})
also have clear peaks. 
However, those peaks appear to be too broad to provide a convincing explanation for the excess of the leading two jet transverse 
energy distributions at CDF~\cite{CDF-bbb-url}. We return to this issue in Section 4.2.

To study the octet discovery limit at the Tevatron, we impose high $p_T$ cuts on the three leading $b$ jets, of 150, 120, and 100 GeV, respectively.  For $M_{G_H} = 300$~GeV, the signal is simulated to have an acceptance around 3\%. Further imposing the invariant mass cut  $|m(b_1, b_2)-M_{G_H}| < 0.2 M_{G_H}$, there are approximate 6 signal events for 20 fb$^{-1}$ with approximately zero background. Therefore, using only the three leading $b$ jets, we estimate that the Tevatron can discover this octet particle up to a mass of $\sim 300$~GeV.   

\subsection{Effects of boundstates}

Since the octet is much heavier than the QCD confinement scale, the QCD interaction approximately generates an attractive Coulomb-like potential between two octets. We anticipate various boundstates named octonium from two octets. From group theory, there are only three attractive channels: $1$, $8_{\rm A}$ (antisymmetric) and  $8_{\rm S}$ (symmetric) from $8 \otimes 8$, because they have positive differences of the quadratic Casimir between the initial and the final states. Defining $C_f = [C_2(8) + C_2(8) - C_2(f)]/2$, we have $C_1 = 3$ and $C_{8_{\rm A}}=C_{8_{\rm S}}=3/2$. The potential from the one-gluon exchange is 
\be
V(r) \,=\, -\,C_f\,\frac{\alpha_s(1/r)}{r}\,,
\label{eq:potential}
\ee
where $\alpha_s$ is evaluated at the energy scale of the Bohr radius of the boundstate. In this paper, we will evaluate the QCD coupling running only at one loop. The gauge coupling in terms of the Bohr radius is given by
\be
\alpha_s(1/r) \,=\,\frac{1}{\alpha_s^{-1}(M_Z) -  b_0 \ln\left( r M_Z  \right)}\,.
\ee 
Here, $b_0 = (11 N_c - 2 N_f)/(6\pi)$ with $N_f=5$ for $m_b < 1/r < m_t$. 

We solve the Schr$\ddot{\rm o}$dinger equation numerically for the potential (\ref{eq:potential}) using the above running gauge coupling.
The Bohr radius is  $r_0 \approx 0.07$~GeV$^{-1}$, the binding energy is $E_b(8) \approx 2.0$~GeV and the wavefunction squared at the origin is $|\psi_8(0)|^2 \approx (9.2~\mbox{GeV})^3$. To compare with the pure Coulomb potential, we fix the gauge coupling to be $\alpha_s(1/r_0)$
and find $E_b = C_f^2\,\alpha_s^2(1/r_0)\,M_{G_H}/4=2.2$~GeV and $|\psi(0)|^2 = C_f^3\,\alpha_s^3(1/r_0)\,M_{G_H}^3/(8\pi) \approx (12.4~\mbox{GeV})^3$. So, the effects of these two potentials differ by about 10\%. In following, we will use  the more precise running potential. We also report the binding energy and wavefunction squared for the singlet channel as $E_b(1) \approx 6.5$~GeV and $|\psi_1(0)|^2 \approx (16.9~\mbox{GeV})^3$.

In our calculation of the boundstate wavefunction and the binding energy, we have neglected the width of the constituents. If the octet width is much larger than the binding energy [as is the case for $m_\psi/|\lambda_3 \eta_3|\sim 100$~GeV in Eq.~(\ref{eq:decaybb})], the boundstate's effects on the octet production can be neglected. 

For a narrower octet, with $\Gamma (G_H) \ll E_b$, we neglect the width of the constituents.  The $S$-wave boundstates of two scalar fields have the quantum numbers, $J^{PC}$, of $0^{++}$ for the singlet, $0^{+-}$ for $8_{\rm A}$ and $0^{++}$ for $8_{\rm S}$. Their couplings to the quarks should be suppressed by the quark mass because of the chirality, so the dominant production cross section should be from gluons at colliders. Since the production cross section is related to the decaying widths via the Breit-Wigner formula, we first calculate the decaying widths of those three boundstates to $gg$ and $2b+2\bar{b}$. The falling apart width is $\Gamma_{4b} \approx 2\,\Gamma_G$. The  decay widths of the boundstate ${\cal B}_f$ to two gluons are calculated using the non-relativistic relation at the low-velocity limit, $\Gamma({\cal B}_f \rightarrow X) = v \sigma(G_H G_H \rightarrow X) |\psi(0)|^2$, as~\cite{Cheung:2004ad}
\bea
\Gamma(1 \rightarrow gg) \,=\,\frac{9\,\pi\,\alpha_s^2(M_{{\cal B}_1})}{2\,M^2_G}\,|\psi_1(0)|^2\,, \qquad \Gamma(8_{\rm S} \rightarrow gg) \,=\,\frac{9\,\pi\,\alpha_s^2(M_{{\cal B}_8})}{8\,M^2_G}\,|\psi_{8_{\rm S}}(0)|^2\,,
\eea
where the mass of the boundstate is $M_{{\cal B}_f} = 2 M_{G_H} - E_b(f)$. Note that $8_A$ does not decay into two gluons.
For $M_{G_H} = 150$~GeV, we have $\Gamma(1 \rightarrow gg)\approx 0.04$~GeV and $\Gamma(8_{\rm S} \rightarrow gg)\approx 0.002$~GeV.

In the narrow width approximation, we have the following partonic production cross section of those boundstates
\bea
\hat{\sigma}_{gg \rightarrow 1} (\hat{s}) = \frac{\pi^2\,\Gamma(1 \rightarrow gg)}{ 8\,M_{{\cal B}_1} }\,\delta(\hat{s} \,-\,M^2_{{\cal B}_1}  )\,, \qquad 
\hat{\sigma}_{gg \rightarrow 8_{\rm S}} (\hat{s}) = \frac{\pi^2\,\Gamma(8_{\rm S} \rightarrow gg)}{M_{{\cal B}_8} }\,\delta(\hat{s} \,-\,M^2_{{\cal B}_8}  )\,.
\eea
Using the gluon PDF, we obtain the production cross section at the Tevatron with $\sqrt{s} = E_{\rm CM} = 1.96$~TeV 
\bea
\sigma(p\bar{p}\rightarrow {\cal B}_f) = \frac{\pi^2\,\Gamma(1 \rightarrow gg)}{8\,M_{{\cal B}_f}\,s}\,\int^1_{M_{{\cal B}_f}^2/s} \frac{dx}{x}\,f_g(x)\,f_g\left(\frac{M^2_{{\cal B}_f}}{x\,s}\right)\,.
\eea
For $M_{G_H} = 150$~GeV, using the Mathematica MSTW 2008 PDFs \cite{Martin:2009iq}, the resonance production cross sections are $\sigma(p\bar{p}\rightarrow 1) \approx 190$~fb and $\sigma(pp\rightarrow 8_{\rm S}) \approx 57$~fb, which is a few percent of the octet pair production cross section and can be neglected in the collider searches. 


\subsection{Supersymmetric Higgs bosons}\label{sec:SUSYHiggs}

Let us now compare the kinematic distributions of the multi-$b$-jet final states arising from scalar octet production with those due to 
the Higgs bosons of the MSSM.
In Two-Higgs-doublet models, such as the Higgs sector of the MSSM, there are two color-singlet and electrically-neutral spin-0 particles 
(the heavy Higgs bosons $H^0$ and $A^0$) 
that may have large couplings to the $b$ quark. In the 5-flavor PDF scheme, where the proton includes a $b$-quark, 
the emission of  a heavy Higgs boson from a $b$ quark line leads to a $3b$ final state (see diagram next to Table 2).
Equivalently, in the 4-flavor PDF scheme, where there is no $b$ inside the proton, production of a $b\bar{b}$ pair followed by emission of
a heavy Higgs boson gives four $b$ jets that may appear as a $3b$ final state when one of the jets does not pass the cuts.
At large $\tan \beta$ and for $M_A \gg M_Z$, the production cross sections of  $H^0$ and  $A^0$ are approximately equal, and are related to the standard model  Higgs production by $\sigma (b\bar{b}H^0)=\sigma (b\bar{b}A^0) = \tan^{2}\!\beta \, \sigma(b\bar{b}h^0_{\rm SM})$ \cite{Djouadi:2005gj}. 

\begin{figure}[t]
\begin{center}
\vspace*{3mm}
\includegraphics[width=0.48\textwidth]{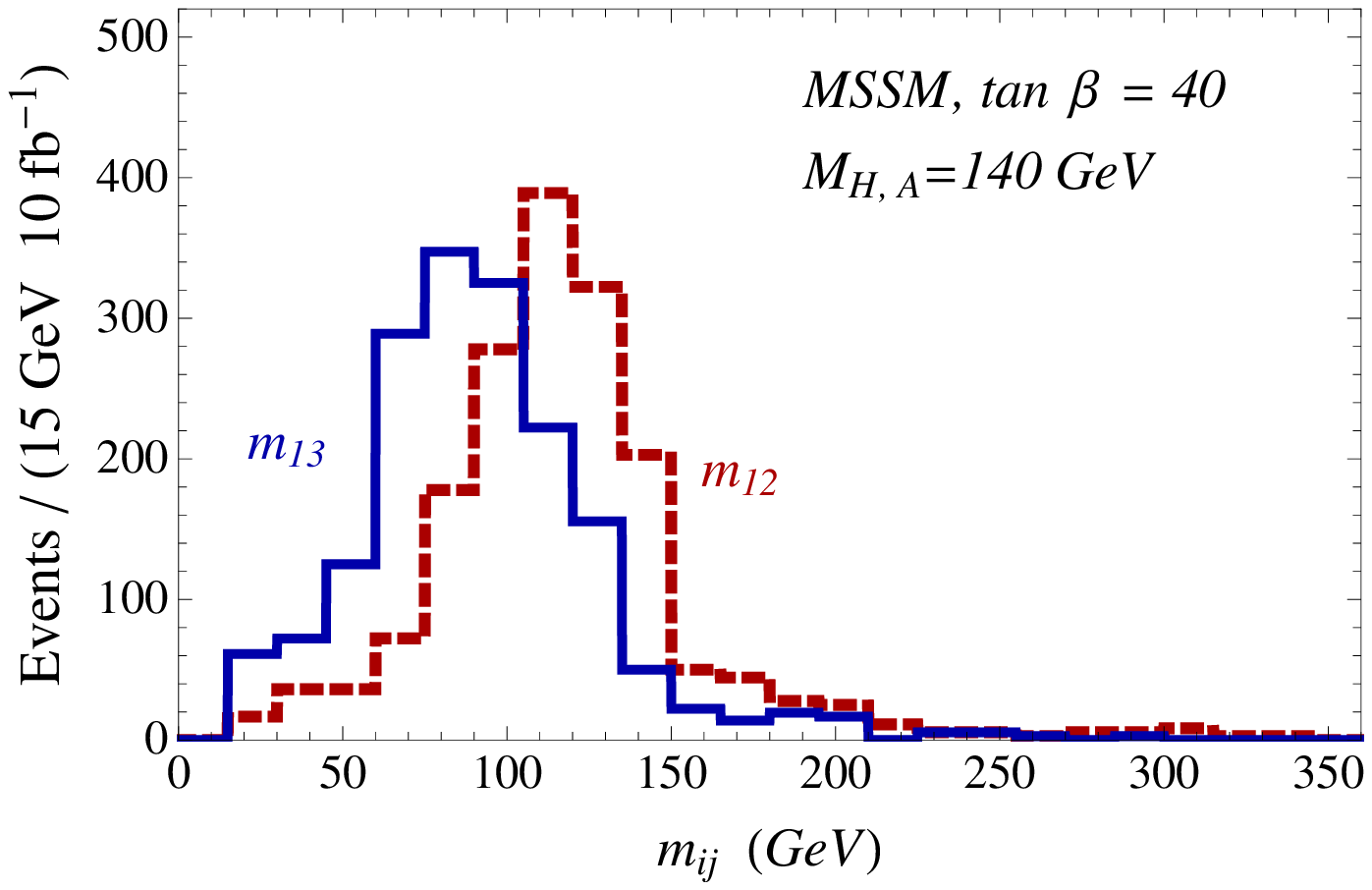} \hspace{2mm}
\includegraphics[width=0.48\textwidth]{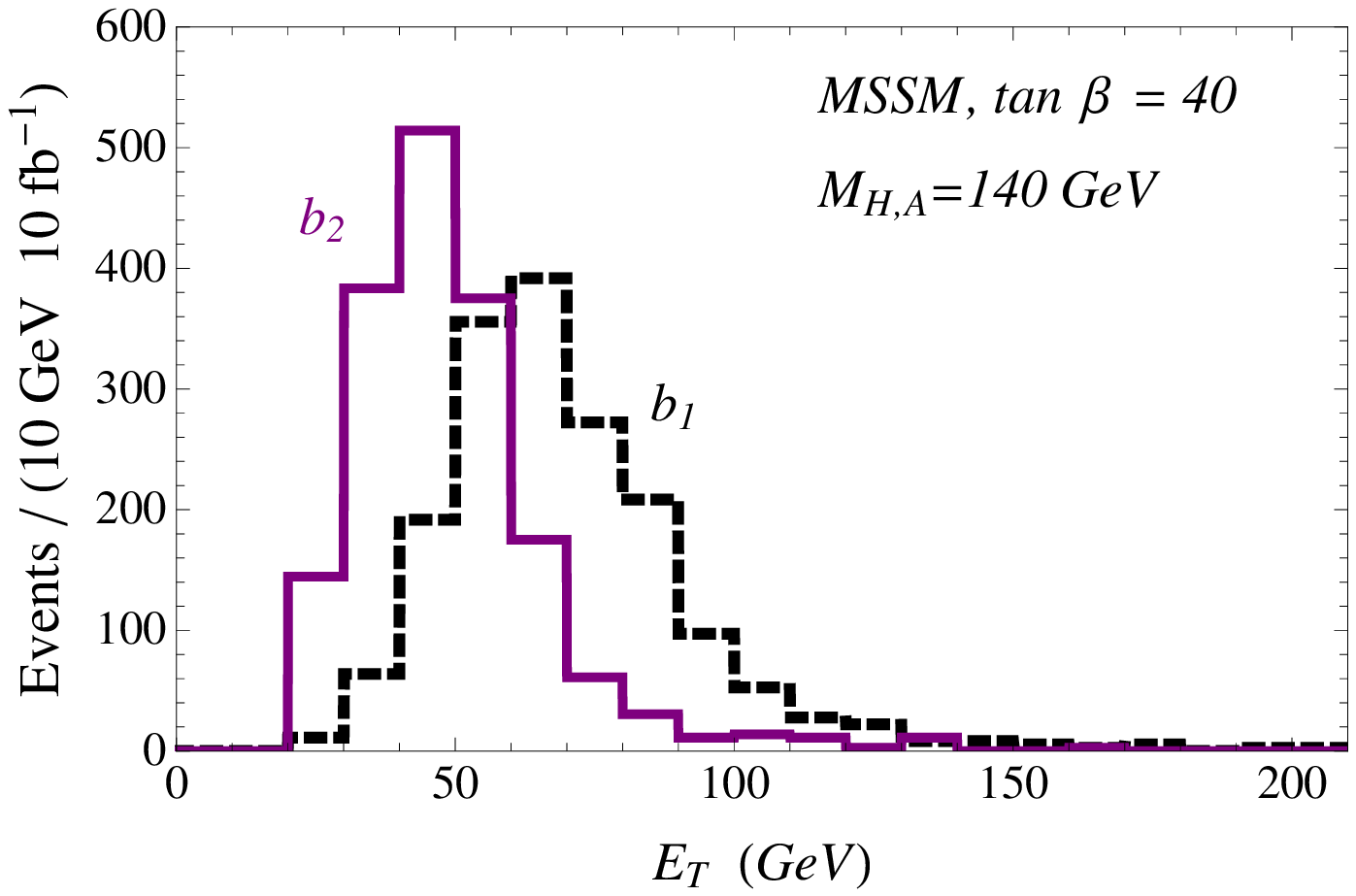}
\caption{Same as Fig.~\ref{fig:mijEt}, but for MSSM Higgs production at $\tan\beta = 40$ (the leading-order production cross section is around 5.4~pb).}
\label{fig:SusymijEt}
\end{center}
\end{figure}

In order to study the multi-$b$ Higgs signal, we impose the same cuts as in the case of scalar octets: $p_T > 20$~GeV and $|\eta| < 2$ for 
each $b$ jet. Compared to the octet case, fewer events pass these cuts because one of the jets does 
not arise from the decay of a heavy particle.
For example, the fourth $b$ jet has $\sim$3 times smaller chance to pass the basic cuts than in the octet case (see rows 2 and 3 of Table~\ref{tab:SUSYcomp}). 
The invariant mass of the leading and the third jets is also less likely to be within the mass window $|m_{ij}-M_{G_H}| < 0.2 M_{G_H}$ than 
in the octet case (see last two rows of Table~\ref{tab:SUSYcomp}). Thus, the $m_{13}$ distribution is a good discriminant between 
the octet and the MSSM Higgs bosons, especially when the resonance is heavy. 
We show the invariant mass distributions for the  $b_1, b_2$ and $b_1, b_3$ jet pair in Fig.~\ref{fig:SusymijEt} for heavy Higgs bosons of mass 140 GeV
and a leading-order production cross section (summed over $A^0$ and $H^0$) corresponding to $\tan\beta=40$. 
Comparing Figs.~\ref{fig:SusymijEt} and \ref{fig:mijEt}, we can see that the MSSM Higgs bosons lead to an $m_{13}$ distribution with a peak that is
both lower and located at a smaller value than the peak due to the scalar octet.

\begin{table}[t!]
\label{tab:susyhiggs}
\vspace*{-20mm}
\renewcommand{\arraystretch}{1.5}
\hspace*{-11mm}
\centerline{
\begin{tabular}{|c|ccccc|}
\hline
$M$ (GeV)&150 & 200 &250 &300 &350 \\ \hline\hline
$G_H$  & 15.8  & 16.6  & 19.7 & 19.1  & 21.0          \\ \hline
$H^0, A^0$ & 4.3  & 6.1  & 7.4 & 7.7  & 7.7   \\ \hline \hline
$G_H$   & 1.0  & 1.0  & 0.86 & 0.79  & 0.73          \\ \hline
$H^0, A^0$  & 0.50  & 0.25  & 0.16 & 0.10  & 0.10   \\ \hline 
\end{tabular}
\unitlength=1.2 pt
\SetScale{1.2}
\SetWidth{1}      
\normalsize    
{} \allowbreak
\begin{picture}(100,100)(-60,30)
\ArrowLine(10,50)(45,70)\ArrowLine(10,20)(10,50)
\ArrowLine(-30,0)(10,20)
\Gluon(10,50)(-30,70){-3}{6}
\DashLine(10,20)(45,15){3}
\ArrowLine(45,15)(80,25)\ArrowLine(45,15)(80,5)
\Text(88,5)[c]{\small $b$}\Text(88,25)[c]{\small $\bar{b}$}
\Text(47,62)[c]{\small $b$}\Text(16,38)[c]{\small $b$}
\Text( -34,59)[c]{\small $g$}
\Text( -34,8)[c]{\small $b$}
\rotatebox{-9}{\Text(18,12)[c]{\small $H^0$, $A^0$}}
\end{picture}
}
\vspace*{3mm}
\caption{Ratio (in \%) of the numbers of events with $4b$ or $3b$, for the MSSM (representative process shown on the right-hand side) 
at $\tan\beta = 40$ (3rd row)
and QCD $G_HG_H$ production (2nd row). Also, the ratio of numbers of events that pass the $m_{12}$ or $m_{13}$ invariant-mass window cut, $|m_{ij}-M_{G_H}| < 0.2 M_{G_H}$,
 for the MSSM (5th row) and for the octet (4th row). 
}
\label{tab:SUSYcomp}
\end{table}

If a large excess of events over the standard model background will be established in the $3b$ final state, then an MSSM interpretation will 
typically require an excess of events
in the $b\tau^+\tau^-$ final state due to the decays of $H^0$ and $A^0$ into tau leptons. By contrast, $G_H$ can not decay into leptons-only 
final states because it is a color octet, so that no $b\tau^+\tau^-$ excess is predicted in the octet models.

\section{Renormalizable Coloron Model}\label{sec:ReCoM}\setcounter{equation}{0}

Let us now analyze a minimal renormalizable model that includes a spin-1 color-octet particle \cite{Hill:1991at, Chivukula:1996yr}, called ``coloron''.
We will show that this model includes a scalar octet, identical with the $G_H$ studied in sections \ref{sec:color-octet} and \ref{sec:Tevatron} except that here it can be resonantly produced in pairs, through an $s$-channel coloron, $G_\mu^\prime$.
The gauge symmetry of this model is an extension of QCD, $SU(3)_1 \times SU(3)_2$. This symmetry is spontaneously broken down to 
the QCD gauge group $SU(3)_c$ by the VEV of a complex scalar field, $\Sigma$, which transforms as $(3, \bar{3})$ under $SU(3)_1 \times SU(3)_2$.

\subsection{Interactions and masses}

The most general renormalizable potential of $\Sigma$ is
\be
V(\Sigma) \,=\, -m_\Sigma^2\, \Tr(\Sigma \Sigma^\dagger) \, -\, \mu \left( {\rm det} \, \Sigma + {\rm H.c.} \right)  +
\frac{\lambda}{2}\left[ \Tr \left(\Sigma \Sigma^\dagger\right)  \right]^2 \,+\,\frac{\kappa}{2}\,\Tr \left(\Sigma \Sigma^\dagger \Sigma \Sigma^\dagger \right) ~~,
\label{eq:sigma-pot}
\ee
where without loss of generality we take $\mu >0$. Note that the second term above is $SU(3)_1 \times SU(3)_2$ invariant because it can be written as
\be
{\rm det} \, \Sigma = \frac{1}{6} \epsilon^{ijk}\,\epsilon^{i^\prime j^\prime k^\prime}\,\Sigma_{i i^\prime}\,\Sigma_{j j^\prime}\,\Sigma_{k k^\prime}  ~~.
\ee
We assume $m_\Sigma^2 >0$ so that  $\Sigma$ acquires a VEV:
\be
\langle \Sigma \rangle \, = \,\frac{f_{\Sigma}}{\sqrt{6}}\, \mathbb{I}_3 \,=
 \frac{ \sqrt{4(\kappa + 3\,\lambda)\,m_\Sigma^2 + \mu^2 }  + \mu   }{2(\kappa + 3\,\lambda )} \; \mathbb{I}_3  ~~,
 \label{eq:vev}
\ee
where $\mathbb{I}_3$ is the unit $3\times 3$ matrix.
The potential is bounded from below provided $3\lambda+\kappa > 0$.

Expanding  $\Sigma$ around this vacuum, we find that its 18 degrees of freedom ($\Sigma$ is a $3\times 3$ complex matrix)
are grouped into four real scalar fields: two octets ($G_H^a$ and $G_G^a$, $a = 1,..., 8$)
and two singlets ($\phi_R$ and $\phi_I$) under $SU(3)_c$,
\be
\Sigma = \frac{1}{\sqrt{6}}(f_\Sigma + \phi_R + i\,\phi_I)\,\mathbb{I}_3 \, +\, \left( G_H^a \, + \, i\,G_G^a \right) T^a ~~.
\label{eq:sigmapara}
\ee
$G_G^a$ are the Nambu-Goldstone bosons associated with the broken generators of $SU(3)_1 \times SU(3)_2$, and become the longitudinal 
degrees of freedom of the coloron.
Previous studies of this model have assumed that the other degrees of freedom of $\Sigma$ are heavy enough to be 
neglected\footnote{An exception is the $SU(3)_1 \times SU(3)_2$  linear $\sigma$-model description of heavy-light mesons in QCD \cite{Bardeen:1993ae}.}.
It turns out, however, that within a renormalizable model valid for a range of scales above $f_{\Sigma}$, the $G_H^a$, $\phi_R$ and $\phi_I$
scalars are lighter than the coloron. Thus, it is interesting to study their properties. We refer to the renormalizable model that includes all the scalar
particles contained in $\Sigma$ and the $SU(3)_1 \times SU(3)_2$ gauge symmetry as ReCoM.

For $\mu \rightarrow 0$, there is a $U(1)_1\times U(1)_2$ global symmetry broken by $\langle \Sigma \rangle$ down to  
the diagonal $U(1)$ subgroup, and 
the associated Nambu-Goldstone boson is $\phi_I$, which becomes massless. For any $\mu$, the squared mass of  $\phi_I$
is given by 
\be
M^2_{\phi_I}  = \sqrt{\frac{3}{2}}\, \mu \,f_\Sigma  ~~.
\label{phi-mass}
\ee
For $\kappa \rightarrow 0$ and $\mu \rightarrow 0$ the potential (\ref{eq:sigma-pot}) has a global $SO(18)$ symmetry, which is spontaneously broken 
by the $\Sigma$ VEV down to $SO(17)$, so that $G_H$ and $\phi_I$ are massless Nambu-Goldstone bosons. 
Hence, the quartic term in $V(\Sigma)$
proportional to $\lambda$ does not contribute to the $G_H$ mass. For any $\kappa$ and $\mu$, the squared mass of $G_H$ is 
\be
M^2_{G_H} = \frac{1}{3} \left( 2 M^2_{\phi_I} + \kappa\,f_\Sigma^2 \right) ~~.
\label{GH-mass}
\ee
The mass of  $\phi_R$ also follows from Eq.~(\ref{eq:sigma-pot}):
\be
M^2_{\phi_R} = \frac{1}{3} \left[\left(\kappa + 3\lambda \right)f_\Sigma^2  - M^2_{\phi_I} \right]  ~~.
\ee
Eq.~(\ref{eq:vev}) implies $f_\Sigma \geq \sqrt{6}\,\mu/(\kappa + 3\lambda)$, so that $M^2_{\phi_R} \geq 0$. 

The kinetic term of $\Sigma$, normalized such that all its component fields shown in Eq.~(\ref{eq:sigmapara}) 
have canonical kinetic terms, is given by 
\be
{\rm Tr}\left( D_\mu \Sigma^\dagger \, D^\mu \Sigma \right) ~~,
\label{eq:sigmakinetic}
\ee
where the $SU(3)_1 \times SU(3)_2$ covariant derivative is 
\be
D^\mu \Sigma = \partial^\mu \Sigma - i\,h_1\,G_1^{\mu\,a}\,T^a\,\Sigma \,+\,i\,h_2\,G_2^{\mu\,a}\,\Sigma\,T^a ~~.
\ee
The kinetic term leads to a mass-square matrix for the two gauge fields, $G_1^{\mu\,a}$ and $G_2^{\mu\,a}$, proportional to $f_\Sigma^2$.
Upon diagonalization, one linear combination becomes the massless QCD gluon,
\be
G^\mu  = \cos{\theta}\,G_1^\mu + \sin{\theta}\,G_2^\mu ~~,
\ee
while the orthogonal linear combination is a massive spin-1 octet, the coloron:
\be
G^{\prime \mu} = \sin{\theta}\,G_1^\mu - \cos{\theta}\,G_2^\mu  ~~.
\ee
The mixing angle depends only on the $SU(3)_1 \times SU(3)_2$ gauge couplings, $h_1$ and $h_2$:
 $\tan{\theta} = h_1/h_2$. The QCD gauge coupling is given by
$g_s = h_1 \cos{\theta} = h_2 \sin{\theta}$, and the mass of the coloron is
\be
M_{G^\prime} \,=\, \frac{\sqrt{2}\,  g_s}{\sqrt{3} \, \sin{2\theta}}\,f_\Sigma \,.
\ee
This mass is larger than the VEV $f_\Sigma$ if $\tan\theta \lae 0.65$. 
For $\tan\theta \ll 1$ and $\lambda, \kappa \lae O(1)$, the coloron is heavier than the scalars. 
For example, the set of parameters
\be
\tan{\theta} = 0.12  \;\; ,  \;\; f_\Sigma = 75 \; {\rm GeV} \;\; ,  \;\; \mu = 275  \; {\rm GeV} \;\; ,  \;\; \kappa = \lambda = 1.4  \;\; ,
\label{inputs}
\ee
and $\alpha_s(M_{G^\prime}) \simeq 0.1$ gives the following mass spectrum:
\be
M_{G^\prime} \approx 290  \; {\rm GeV} \;\; ,  \;\; M_{G_H} \approx 140 \; {\rm GeV} \;\; ,  \;\; M_{\phi_I} \approx 159  \; {\rm GeV} \;\; ,  \;\;  M_{\phi_R} \approx 46   \; {\rm GeV} \;\; .
\ee  

Substituting the decomposition of $\Sigma$, Eq.~(\ref{eq:sigmapara}), into the kinetic term (\ref{eq:sigmakinetic}), 
we find the following interactions of a coloron and two scalars:
\be
\frac{\sqrt{2}\, g_s}{\sqrt{3}\, \sin{2\theta}} \, G_\mu^{\prime\,a}\left(\phi_I \partial^\mu G_H^a  - G_H^a \partial^\mu \phi_I \right) 
+\frac{g_s}{\tan{2\theta}}\,f^{abc}\, G_\mu^{\prime\,a}\,G_H^b\,\partial^\mu G_H^c ~~.
\ee
These interactions induce $G^\prime$ decays into scalars with tree-level widths (in the case where the two-body decays are kinematically open)
given by:
\bear
&& \hspace*{-2.3cm} \Gamma(G^\prime_\mu \rightarrow G_H\,G_H) =  \frac{\alpha_s}{ 16\,\tan^2{2\theta}}\,M_{G^\prime}
\left( 1\,-\,\frac{4\,M_{G_H}^2}{M^2_{G^\prime}}\right)^{3/2} ~~,
\nonumber \\
&& \hspace*{-2.3cm} \Gamma(G^\prime_\mu \rightarrow G_H\,\phi_I) = 
\frac{\alpha_s}{18\,\sin^2{2\theta}}\,M_{G^\prime} \left( 1 - 2\frac{M_{\phi_I}^2 + M_{G_H}^2}{M_{G^\prime}^2} 
+ \frac{\left( M_{\phi_I}^2 - M_{G_H}^2 \right)^2}{M_{G^\prime}^4} \right)^{\! 3/2}   ~~. 
\label{decays}
\eear

If all standard model quarks belong to the fundamental representation of $SU(3)_1$, then $G^\prime$ couples to them as
\be
g_s \tan{\theta}\;\overline{q}\gamma^\mu T^a G_\mu^{\prime\,a}q  ~~.
\ee
The width of the coloron decay into quark pairs is 
\be
\Gamma(G^\prime_\mu \rightarrow q\,\bar{q}) \,=\, \frac{\alpha_s}{6} \tan^2{\!\theta}\,M_{G^\prime}
\left( 1\,-\,\frac{4\,m_q^2}{M^2_{G^\prime}}\right)^{\! 1/2} ~~,
\label{jj-decays}
\ee
where we have not summed over quark flavors. The kinematic suppression in the last factor is relevant for $G_\mu^\prime \rightarrow t\bar{t}$,
if $M_{G^\prime} > 2 m_t$. The coloron effects on top-quark physics are rather sensitive to new quarks that mix with the top \cite{Dobrescu:2009vz}
and to the top charges under $SU(3)_1\times SU(3)_2$ \cite{Hill:1991at}.

QCD corrections to the above decay widths may be large, perhaps of the order of 50\%, but computing them is beyond the scope of this paper.
Assuming that the QCD corrections to $G^\prime_\mu \rightarrow G_H\,G_H$ and $G^\prime_\mu \rightarrow q\,\bar{q}$ decays are of the same size,
the parameters given in Eq.~(\ref{inputs}) lead to branching fractions of 61\% and 39\% for $G_HG_H$ and $jj$, respectively.

\subsection{ReCoM and the CDF excess in the 3$b$ final state}

\begin{figure}[b!]
\begin{center}
\includegraphics[width=0.58\textwidth]{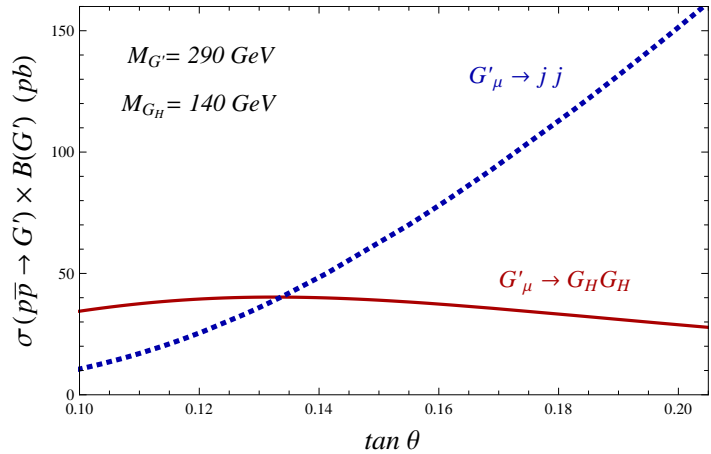}
\caption{Production cross section of $G^\prime_\mu$ at the Tevatron times branching fractions, computed in the narrow width approximation and at leading order. The cross section for  $p\bar{p} \to G^\prime_\mu \to G_H G_H$ is given by the 
solid (red) line, and that for  $p\bar{p} \to G^\prime_\mu \to j j$ by the dotted (blue) line. 
The masses are taken to be  140~GeV for $G_H$ and  290 GeV for $G^\prime_\mu$, and the decay $G^\prime_\mu \rightarrow G_H\,\phi_I$ is assumed to be 
kinematically forbidden. }
\label{fig:Gp-prod}
\end{center}
\end{figure}

The total width of $G_\mu^\prime$ is rather small, typically less than one percent of its mass for $\tan\theta \lae 0.3$ and for a 
$G_\mu^\prime$ mass not more than a few percent above the $G_HG_H$ and $G_H \phi_I$ thresholds.
We can then use the narrow width approximation to estimate the cross section for producing a $G_\mu^\prime$ in the $s$-channel:
\be
\sigma (q\bar{q} \to G^\prime ) \approx \frac{ 8 \pi^2 \alpha_s \tan^2{\!\theta} }{9 M_{G^\prime}}  \;  \delta \left(\sqrt{\hat{s}} - M_{G^\prime}\right)  ~~.
\ee
Convoluting this partonic cross section with the MSTW \cite{Martin:2009iq} PDFs, and then multiplying by 
the branching fractions derived from Eqs.~(\ref{decays}) and (\ref{jj-decays})
we find the total cross sections shown in Fig.~\ref{fig:Gp-prod}. The QCD corrections to these processes are also likely to be sizable, and need to be computed in the future.

The process  $p\bar{p} \to G^\prime_\mu \to j j$ is constrained by the CDF search for dijet resonances \cite{Aaltonen:2008dn}:
for $M_{G^\prime} \approx 290$ GeV, the limit on the cross section times acceptance is about 100 pb. Given that the $G^\prime_\mu$ production is proportional to  $\tan^2\theta$, Fig.~\ref{fig:Gp-prod} indicates that the CDF limit implies $\tan\theta \lae 0.2$ for a $G_H$ mass of 140~GeV.

The decays of $G_H$ into $b\bar{b}$ or $gg$ proceed through higher-dimensional operators, similar to the case discussed in Section~\ref{sec:color-octet}.
For example, a vectorlike down-type quark $\psi$, which transforms as a triplet under $SU(3)_2$, couples to $\Sigma$ and a $b$ quark: 
\be
\eta_3 \Sigma \,\bar{b}_R \, \psi_L  + {\rm H.c.}
\ee 
Integrating out $\psi$ we obtain at tree level the first operator of Eq.~(\ref{derivative-dim-5}), which mediates the $G_H \to b\bar{b}$ decay,
suppressed by two powers of the $\eta_3$ coupling.
A 1-loop diagram as in Fig.~\ref{fig:B} leads to $G_H \to gg$. This decay also occurs at one loop independently of $\psi$, 
as shown in Fig.~\ref{fig:loops}, due to the $ {\rm det} \, \Sigma $ term from Eq.~(\ref{eq:sigma-pot}).

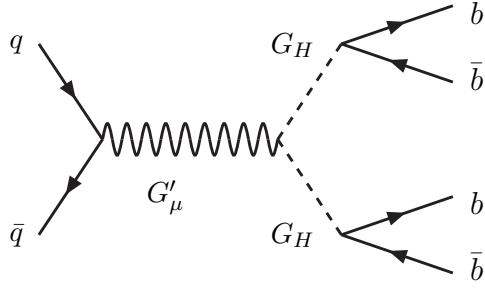
\begin{figure}[t]
\vspace*{-0.3cm}
\begin{center}  \hspace*{-1.1cm}
{
\unitlength=1.2 pt
\SetScale{1.2}
\SetWidth{0.9}      
\normalsize    
{} \allowbreak
\begin{picture}(100,100)(0,0)
\ArrowLine(-10,80)(10,50)
\ArrowLine(10,50)(-10,20)
\Photon(10,50)(65,50){5}{9}
\DashLine(65,50)(85,80){3}
\DashLine(65,50)(85,20){3}
\ArrowLine(85,80)(120,92)\ArrowLine(120,68)(85,80)\ArrowLine(85,20)(120,32)\ArrowLine(120,8)(85,20)
\Text(128,90)[c]{$b$}\Text(128,70)[c]{$\bar{b}$}\Text(128,30)[c]{$b$}\Text(128,10)[c]{$\bar{b}$}
\Text(30,32)[c]{\small $G_\mu^\prime$}
\Text( -17,80)[c]{\small $q$}
\Text( -17,20)[c]{\small $\bar{q}$}
\Text(70,80)[c]{\small $G_H$}
\Text(70,20)[c]{\small $G_H$}
\end{picture}
}
\end{center}
\vspace*{-0.4cm}
\caption{Coloron ($G_\mu^\prime$) resonance decaying to a pair of scalar octets ($G_H$), each 
giving rise to a $b\bar{b}$ resonance.}
\label{fig:ReCoM}
\end{figure}

%
\begin{figure}[t]
\begin{center}
\includegraphics[width=0.48\textwidth]{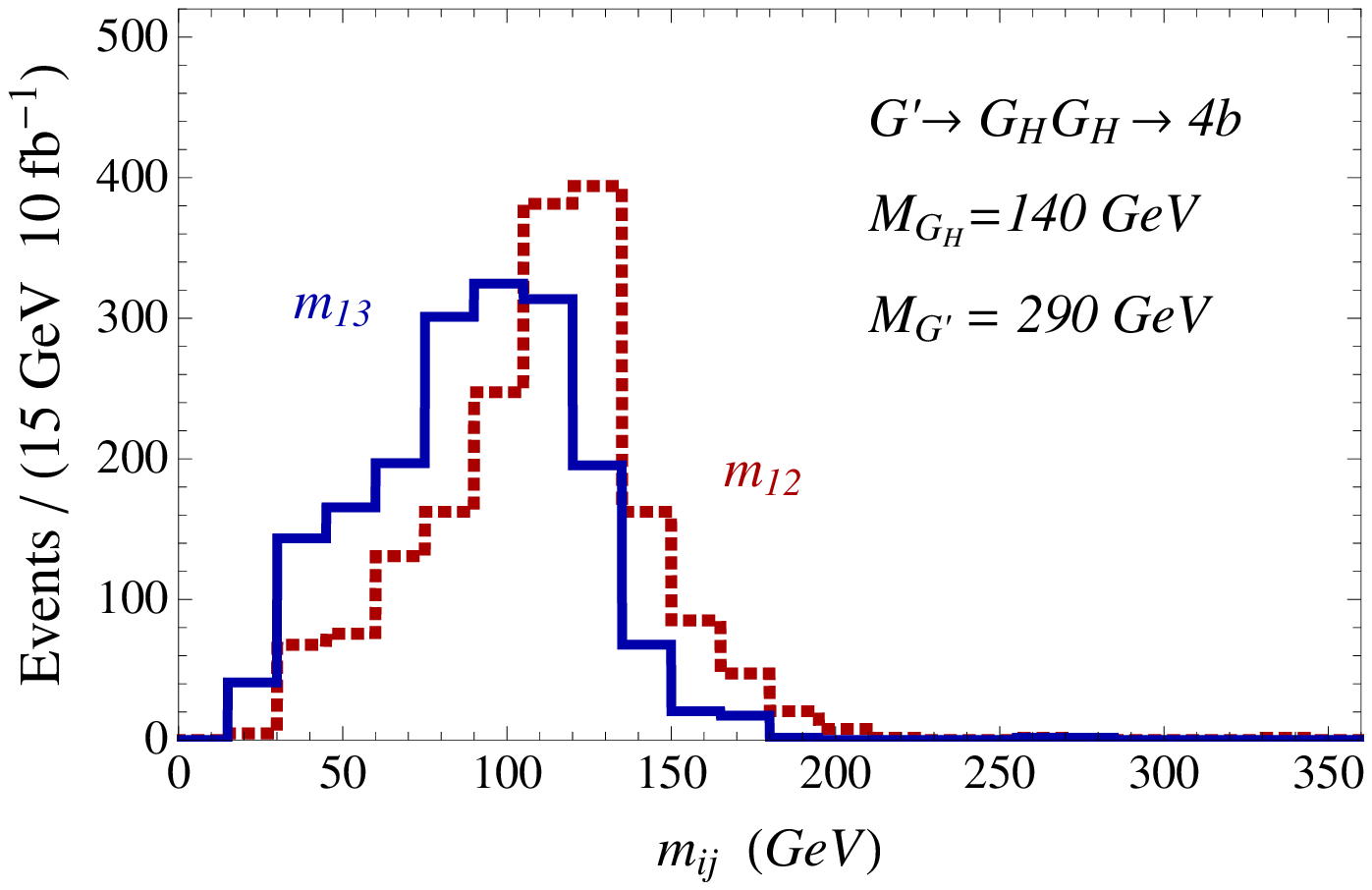} \hspace{2mm}
\includegraphics[width=0.48\textwidth]{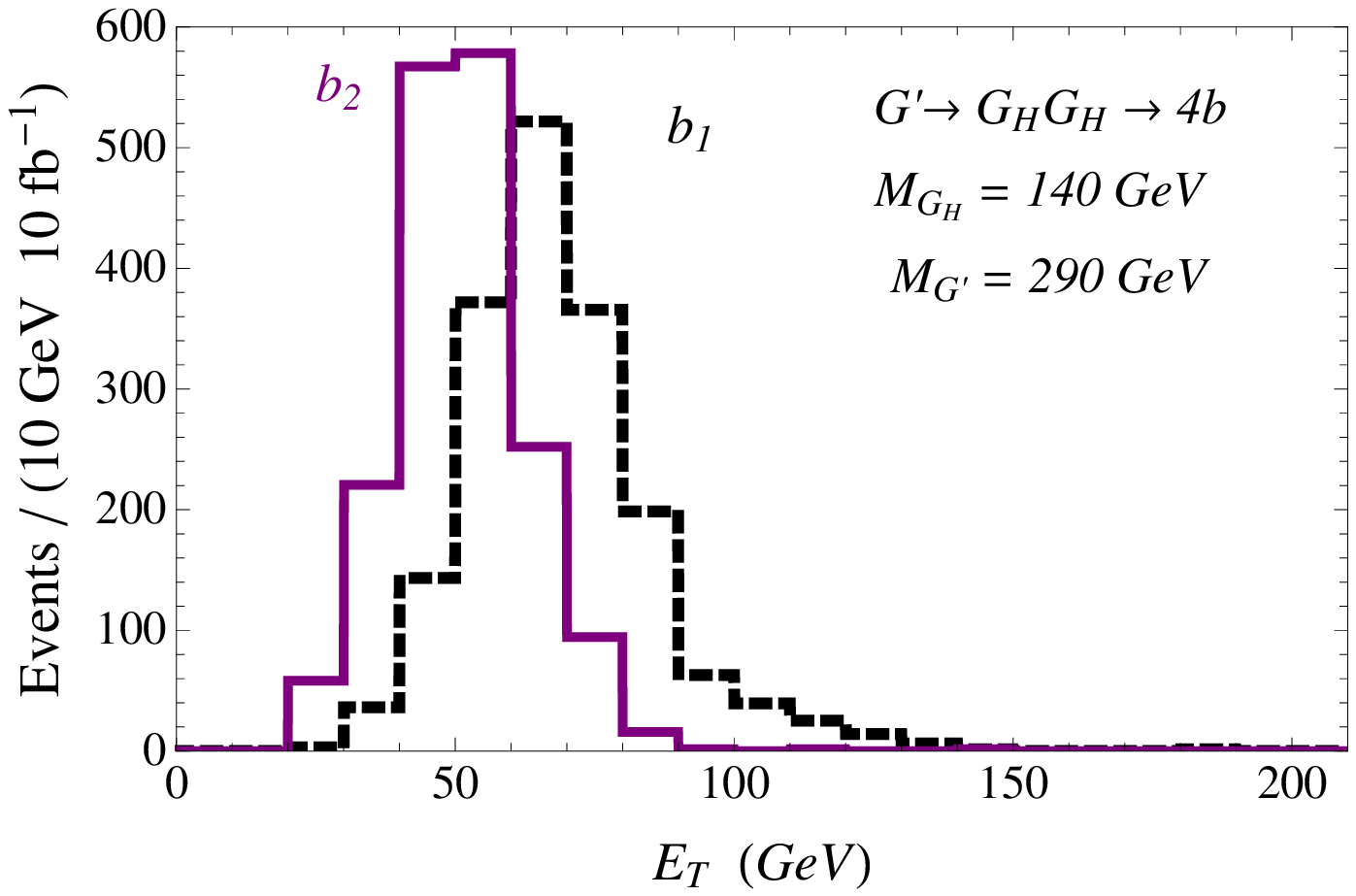}
\caption{Same as Fig.~\ref{fig:mijEt}, but for resonant $G_HG_H$ production via $G^\prime$ and a $G_H \!\to b\bar{b}$ branching fraction of 20\%. The production cross section of  $p\bar{p} \rightarrow G^\prime \rightarrow G_HG_H$ is 40~pb for $M_{G_H} =140$~GeV, $M_{G^\prime}=290$~GeV and $\tan{\theta}=0.12$.}
\label{fig:GpmijEt}
\end{center}
\end{figure}

Compared to the QCD pair production of two $G_H$ scalars, the kinematic distributions of $b$ jets arising from $G^\prime \rightarrow G_HG_H \to 4b$
(see diagram in Fig.~\ref{fig:ReCoM})  
are changed dramatically. Especially when the mass of $G^\prime$ is close to twice of the $G_H$ mass, those two $G_H$'s are mainly produced at rest and the jet $E_T$ distributions are more peaked, as can be seen by comparing the right-hand panels of Figs.~\ref{fig:mijEt} and \ref{fig:GpmijEt}. 
As shown in Fig.~\ref{fig:Gp-prod}, the resonant production of a $G_H$ pair is an order of magnitude larger than the QCD $G_HG_H$ cross section
(of about 4.6~pb for $M_{G_H} = 140$~GeV, as can be seen in Fig.~\ref{fig:cross-section}). Hence, a smaller branching fraction (around 20\%) of $G_H \rightarrow b\bar{b}$ gives a number of $4b$ events comparable to that due to the supersymmetric Higgs bosons for a mass of 140 GeV and $\tan{\beta}=40$.

\begin{figure}[t]
\begin{center}
\hspace*{-2mm}\includegraphics[width=0.495\textwidth]{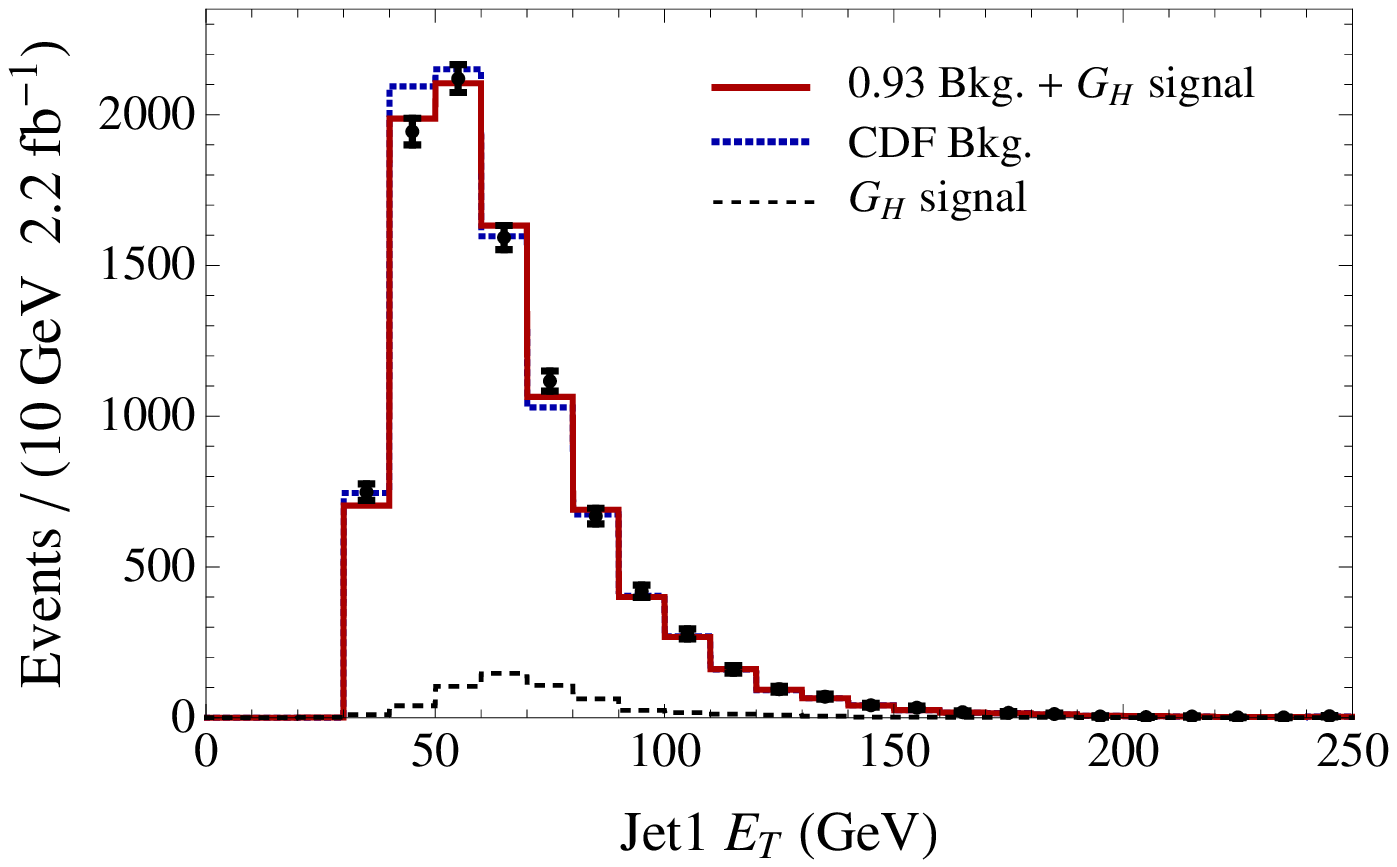} \hspace{.1mm} 
\includegraphics[width=0.495\textwidth]{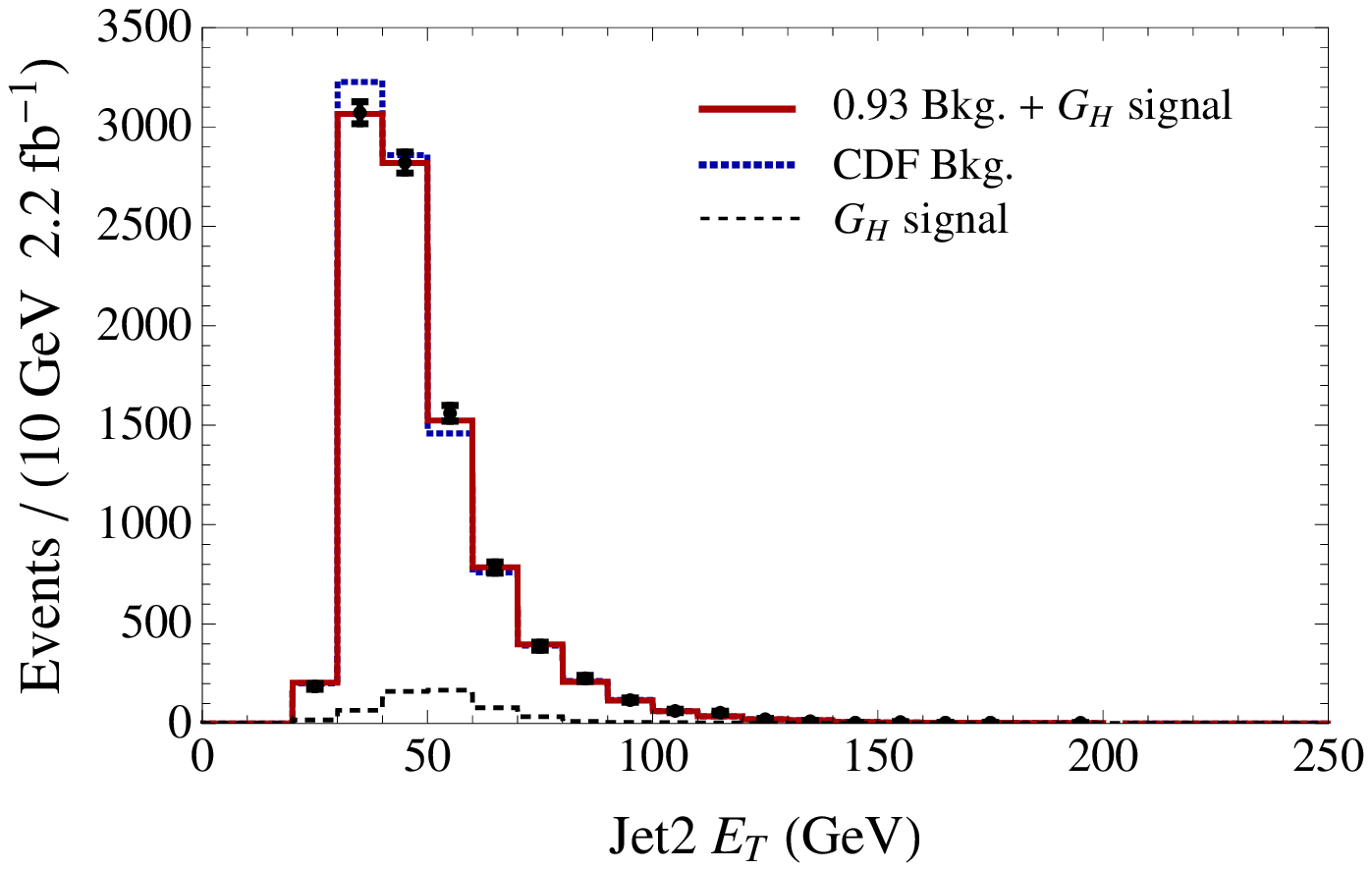} 
\caption{Comparison of the signal-plus-background distributions with CDF data \cite{CDF-bbb-url}. The solid (red) line is a combination of the background (multiplied by 0.93) from CDF and both the resonant (via a $G_\mu^\prime$) and QCD production of two $G_H$ scalars. The branching fraction of $G_H \rightarrow b \bar{b}$ is taken  to be 22\% as a result of fitting to the $m_{12}$ and $E_T$ distributions of the two leading jets. 
The dotted (blue) line, taken from~\cite{CDF-bbb-url}, is the CDF background-only fit to the data. The black points are CDF data with only statistic error bars. The  $G_\mu^\prime$ mass is chosen to be 290~GeV and the $G_H$ mass is 140~GeV.
}
\label{fig:CDF-bbb-url}
\end{center}
\end{figure}

The CDF $3b$ search  \cite{HiggsPlusb} shows an excess in the $m_{12}$ distribution, as mentioned in Section 3.2, and in addition the shapes of the $E_T$ 
distributions of the $b$ jets are shifted towards larger values compared to the standard model background  \cite{CDF-bbb-url}.
The dotted (blue) line in Fig.~\ref{fig:CDF-bbb-url} shows the background from the best 
CDF fit to the data, and the shown data points include only statistical errors. The $40-50$ GeV bin of the leading-jet
$E_T$ distribution is about $3.5\sigma$ below the background, suggesting that the background normalization is too high.
The $70-80$ GeV bin is about $2.5\sigma$ above the potentially overestimated background, suggesting the presence of some 
new heavy particles which produce hard $b$ jets. 
The $E_T$ distribution of the second jet has similar features.

In our ReCoM  there are four parameters relevant for fitting the CDF invariant mass and jet $E_T$ distributions: 
the $G_\mu^\prime$ and $G_H$ masses, the mixing angle $\theta$ and the branching fraction of $G_H \to b \bar{b}$. 
Since the peaks of the $E_T$ distributions of the two leading $b$ jets match the excess bins ($70-80$~GeV for the first jet 
and $50-60$~GeV for the second jet, see Fig.~\ref{fig:CDF-bbb-url}), we fix the masses of $G_\mu^\prime$ and $G_H$ to 
be 290~GeV and 140~GeV, respectively (nearby values with similar splitting, such as 305 GeV and 150 GeV, work equally well). 
Fig.~\ref{fig:Gp-prod} shows that the production cross section is insensitive to $\tan{\theta}$ for $0.1 < \tan{\theta} < 0.2$; 
for illustration, we fix $\tan{\theta} = 0.12$. Taking into account the unknown overall normalization of the background,
we fit the $m_{12}$ (first 10 bins) and $E_T$ (first 9 bins for each of the two leading jets)  distributions reported by CDF \cite{CDF-bbb-url}. 
The total $\chi^2$ reaches the minimum for the background normalization factor of 0.93 and $B(G_H \rightarrow b\bar{b})=22\%$. The $\chi^2$ per degree of freedom is 0.75 ($p$-value of 82\%) from our 
signal-plus-background fit,  while the background-only fit has $\chi^2/dof =1.74$ ($p$-value of 1\%). 
In our simple fit we have not included systematic errors, and we have not allowed different normalization factors for different components of the background. 
Nevertheless, this successful fit indicates that it would be important for the CDF and D0 Collaborations to compare their multi-$b$ data with 
the ReCoM predictions. 

The fitted $G_H \rightarrow b\bar{b}$ branching fraction implies that the $G_H \rightarrow gg$ branching fraction is large,  around $78\%$. Thus,
ReCoM predicts that  the number of $bbjj$ events with equal $b\bar{b}$ and $jj$ invariant masses is about 7 times larger than
the number of $4b$ events satisfying the similar equal-mass condition. 
Another prediction is that the invariant mass of the four jets has a peak close to the coloron mass of 290~GeV.

\subsection{Dijet resonances plus a $W$ boson and other signals} 

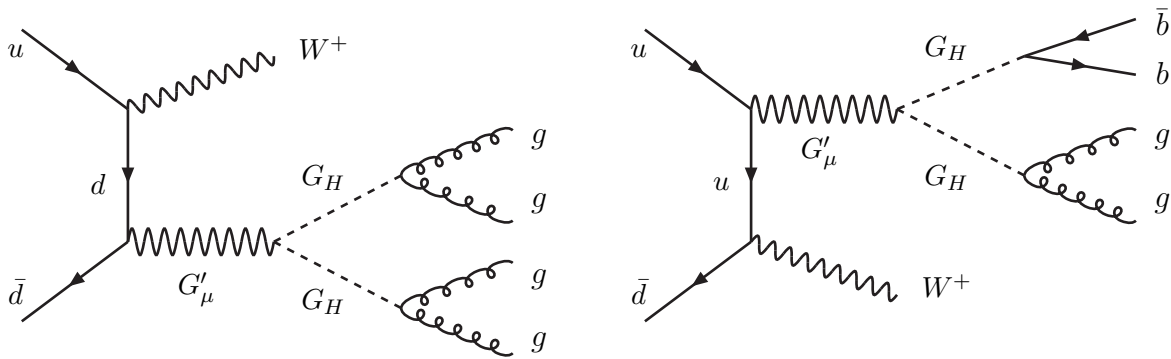
\begin{figure}[b!]
\vspace*{0.4cm}
\begin{center}  \hspace*{-5.8cm}
{
\unitlength=1.2 pt
\SetScale{1.}
\SetWidth{1.}      
\normalsize    
{} \allowbreak
\begin{picture}(100,100)(-26,0)
\ArrowLine(-30,130)(10,100)
\ArrowLine(10,100)(10,50)
\Photon(10,100)(65,120){3}{9}
\ArrowLine(10,50)(-30,20)
\Photon(10,50)(65,50){5}{9}
\DashLine(65,50)(113,75){3}\Gluon(113,75)(155,92){2.2}{5}\Gluon(113,75)(155,58){-2.2}{4}
\DashLine(65,50)(113,25){3}\Gluon(113,25)(155,42){2.2}{4}\Gluon(113,25)(155,8){-2.2}{5}
\Text(138,54)[c]{$g$}\Text(138,74)[c]{$g$}\Text(138,30)[c]{$g$}\Text(138,10)[c]{$g$}
\Text(30,27)[c]{\small $G_\mu^\prime$}
\Text(-27,102)[c]{\small $u$}\Text(-1,60)[c]{\small $d$}\Text(-27,26)[c]{\small $\bar{d}$}
\Text(70,103)[c]{\small $W^+$}
\Text(70,23)[c]{\small $G_H$}\Text(70,62)[c]{\small $G_H$}
\end{picture}
\quad\quad\quad\quad
%
\begin{picture}(100,100)(-80,0)
\ArrowLine(-30,130)(10,100)
\ArrowLine(10,100)(10,50)
\Photon(10,100)(65,100){5}{9}
\ArrowLine(10,50)(-30,20)
\Photon(10,50)(65,30){3}{9}
\DashLine(65,100)(113,120){3}\ArrowLine(155,134)(113,120)\ArrowLine(113,120)(155,113)
\DashLine(65,100)(113,75){3}\Gluon(113,75)(155,92){2.2}{4}\Gluon(113,75)(155,58){-2.2}{5}
\Text(138,54)[c]{$g$}\Text(138,74)[c]{$g$}\Text(138,95)[c]{$b$}\Text(138,111)[c]{$\bar{b}$}
\Text(30,70)[c]{\small $G_\mu^\prime$}
\Text(-27,102)[c]{\small $u$}\Text(-1,60)[c]{\small $u$}\Text(-27,26)[c]{\small $\bar{d}$}
\Text(70,28)[c]{\small $W^+$}
\Text(70,103)[c]{\small $G_H$}\Text(70,62)[c]{\small $G_H$}
\end{picture}
}
\end{center}
\vspace*{-0.6cm}
\caption{Representative diagrams for $W$ boson production in association with a coloron ($G_\mu^\prime$)  
decaying to a pair of scalar octets ($G_H$), each giving rise to a dijet resonance. 
}
\label{fig:GprimeW}
\end{figure}

\begin{figure}[t]
\begin{center}
\includegraphics[width=0.58\textwidth]{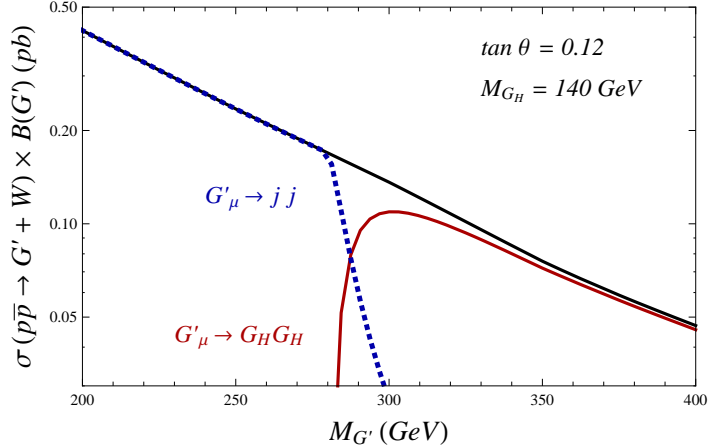}
\caption{Leading-order production cross section (solid black line) of $G^\prime_\mu$ in association with a $W$ boson at the Tevatron
for $\tan{\theta} = 0.12$.
The cross section times the coloron branching fraction is shown 
for the $G^\prime_\mu$ decay into a pair of jets  (dotted line), and into $G_HG_H$ (solid red line).}
\label{fig:GpW-prod}
\end{center}
\end{figure}

Another test of our ReCoM is associated production of $G^\prime_\mu$ with weak gauge bosons. 
Feynman diagrams for producing $G^\prime_\mu$ plus $W^+$ are shown in Fig.~\ref{fig:GprimeW}. We find that the partonic cross section for this process is  
\be
\sigma(u \, \overline{d}\to G_\mu^\prime \,  W^+) =
\frac{8\pi\alpha\alpha_s\tan^2{\theta}}{9\hat{s}^2\sin^2\!\theta_W}
\left[
\frac{\hat{s}^2 \!+\! (M^2_{G^\prime} \! + \! M^2_W)^2 \! }{\hat{s} - M^2_{G^\prime} - M^2_W} 
\ln\! \left( \frac{\hat{s}\!  - \! M^2_{G^\prime} \! -\!  M^2_W \! + \! M_\beta^2  }{2 M_{G^\prime} M_W }  \right) \! - \! M_\beta^2 \right] ~,
\ee
where 
\be
M_\beta^2 = \left[ \left(\hat{s} - M^2_{G^\prime} - M^2_W\right)^2 - 4 M^2_{G^\prime} M^2_W \right]^{1/2} ~.
\ee
The leading-order cross section for the $p\bar{p} \to G_\mu^\prime W$ process at the Tevatron, computed 
using MadGraph, is shown in Fig.~\ref{fig:GpW-prod}
for $\tan{\theta} =0.12$, and is around $150$ fb for $M_{G^\prime} = 290$~GeV. 

%

\begin{figure}[t]
\begin{center}
\includegraphics[width=0.68\textwidth]{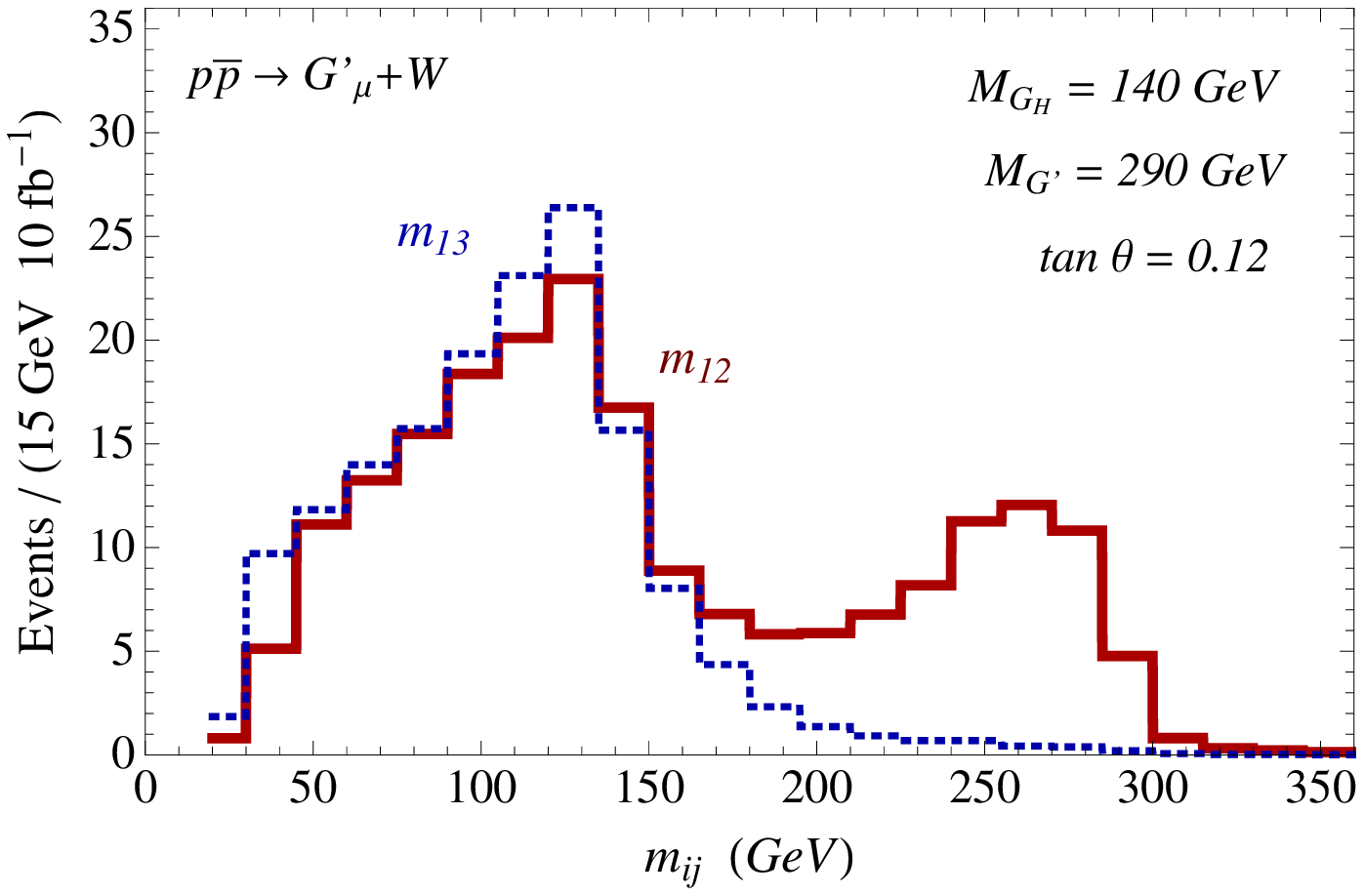} \quad
\caption{Invariant mass distributions for the two leading jets ($m_{12}$, solid line) and for the first and third jet 
($m_{13}$, dashed line),
in events arising from $G^\prime_\mu + W$ production at the Tevatron.
All coloron decay channels and only the $W\to e\nu,\mu\nu$ decays are included.
The first peak is mostly due to the $G^\prime_\mu \rightarrow G_H G_H \rightarrow 4j$ decay (its location 
is sensitive only to the $G_H$ mass), while the second peak in $m_{12}$
is due to the direct $G^\prime_\mu \rightarrow 2j$ decay. The 
branching fractions for those two channels are 61\% and 39\%, respectively, for $M_{G_H} = 140$ GeV, 
$M_{G^\prime} = 290$ GeV and $\tan\theta = 0.12$.
A slightly larger $G^\prime_\mu$ mass increases the first peak, and a larger $\tan\theta$ increases the second peak.} 
\label{fig:GpW-mij}
\end{center}
\end{figure}

Since the coloron can cascade decay into four jets via two $G_H$ scalars, or directly decay into two jets, it is useful to analyze the invariant 
mass distribution of the two leading jets in the final states that include a $W$ decaying to $\ell \nu$, with $\ell = e,\mu$.
Following the CDF search in this channel \cite{CDF-WZ-url}, we impose the $p_T$(jets)$ > 15$~GeV, $p_T(\ell) > 20$~GeV, 
$\slash{\hspace{-2.5mm}E}_T > 25$~GeV cuts on the signal events, using the MadGraph/MadEvents to Pythia to PGS chain as in Section 3. For the signal events from $G_\mu^\prime \rightarrow G_HG_H \rightarrow 4j$, the acceptance for passing those cuts is around 63\% for inclusive 2 jets or 3 jets in the final state. For the signal events from $G_\mu^\prime \rightarrow 2j$, the acceptances are around 59\% for inclusive 2-jet events and 28\% for inclusive 3-jet events. 

In Fig.~\ref{fig:GpW-mij}, we show the invariant mass distributions of a pair of jets in the multi-jet-plus-$\ell\nu$ events resulting from a simulation of the 
ReCoM signal.  As expected, the invariant mass of the two leading jets ($m_{12}$) has two peaks corresponding roughly to the masses of $G_H$ 
and $G^\prime_\mu$, chosen to be 140 GeV and 290 GeV, respectively. The height of the first peak is relatively insensitive to $\tan{\theta}$, since 
$\sigma(p\overline{p} \rightarrow G_\mu^\prime W) \times B(G_\mu^\prime \rightarrow G_H G_H)$ is fairly constant for a range of $\tan{\theta}$ values, which is similar to the case of single $G^\prime_\mu$ production shown by the solid (red) line in Fig.~\ref{fig:Gp-prod}. 
The height of the second peak is strongly sensitive to $\tan{\theta}$; especially for small values of $\tan{\theta}$, the cross section times the 
$B(G_\mu^\prime \rightarrow 2j)$ branching fraction is proportional to $\tan^4\!{\theta}$.

The invariant mass of the leading and third jets ($m_{13}$) has a single peak (see the dashed line in Fig.~\ref{fig:GpW-mij}), near the $G_H$ mass, because 
the direct $G_\mu^\prime \rightarrow 2j$ decay does not contribute at leading order to this distribution. The peak in $m_{13}$ is a distinctive feature of 
ReCoM, allowing to differentiate it 
from the low-scale technicolor model \cite{Eichten:1996dx} that also predicts final states involving a dijet resonance and a $W$ boson.
Even more dramatic would be the observation of two different dijet resonances of equal mass in association with a $W$, as shown in Fig.~\ref{fig:GprimeW},
but in that case it is likely that a computation of next-to-leading order effects is necessary, in order to include the case where an extra jet is radiated with 
a $p_T$ larger than that of the third or fourth jet originating from $G_H$ decays.

$G_\mu^\prime$ production in association with a $Z$ boson or photon may also be interesting. For example, $Z$-plus-jets events at the LHC, with the leading jets forming a resonance and the $Z$ decaying into charged leptons, could allow a sufficient separation of the ReCoM 
signal from the large standard model background. 

Throughout this section we have assumed for simplicity that $\phi_I$ is too heavy to be produced in coloron decays. However, for a range of ReCoM 
parameters [see Eqs.~(\ref{phi-mass}) and (\ref{GH-mass})] $\phi_I$ is  lighter than $G_H$ so that $G^\prime_\mu \to G_H \phi_I$ is the dominant 
decay mode of the coloron. In that case, a larger coloron production cross section ({\it i.e.}, a larger $\tan\theta$) is allowed by the dijet searches because the branching fraction for $G^\prime_\mu \to q\bar{q}$ is suppressed. If  $\phi_I$ decays predominantly into light jets, than the  $G^\prime_\mu \to G_H \phi_I$
decay contributes to the $4j$ and $W+4j$ signals (increasing the height of the first peak in Fig.~\ref{fig:GpW-mij}) without necessarily affecting the multi-$b$ signal. 

\bigskip

\section{Conclusions}\label{sec:conclusion}

Four-jet final states at the Tevatron and the LHC are predicted in various theories 
for physics beyond the standard model.
Even though the QCD background is very large, the presence of invariant mass peaks allows the separation of the signal from the
background \cite{Chivukula:1991zk, Dobrescu:2007yp}.
When the new particles decaying into jets have spin 0, it is natural to expect that a significant fraction of the jets comes from $b$
quarks because the spin-0 couplings to standard model fermions are typically proportional to mass.
Thus, $b$-tagging can further decrease the background.

In this paper we have investigated the properties of weak-singlet, color-octet scalars (``scalar octets'').
Electroweak gauge invariance prevents these scalars from coupling at renormalizable level to standard model fermions,
but dimension-5 couplings to quarks are allowed and lead to decays mostly into $b\bar{b}$ and gluon pairs.
The exchange of a scalar octet of mass below a few hundred GeV could
lead to large CP violation in $B_s$ mixing, as indicated by the D0 dimuon asymmetry (see section 2.3), if a vectorlike 
quark that induces the dimension-5 couplings has mass below the TeV scale.

The scalar octets are pair produced through their coupling to gluons, which is fixed by QCD gauge invariance, or via an $s$-channel resonance due to a spin-1 color-octet particle. We have shown that the simplest gauge invariant origin
of such a resonance, namely the renormalizable version (``ReCoM") of the coloron model based on the $SU(3)_1 \times SU(3)_2$
gauge extension of QCD \cite{Hill:1991at, Chivukula:1996yr},
automatically includes a large coupling of a coloron to a pair of scalar octets. The parameters in this model that affect the
$4b$ signal are the ratio $\tan\theta$ of gauge couplings, which controls both the production and branching fractions of the coloron,
the masses of the coloron and scalar octet, and the branching fraction of the scalar octet into $b\bar{b}$.
We have focused on the case where only three $b$ jets pass some basic cuts, so that a large part of the signal is preserved,
at the expense of not being able to reduce the background as efficiently as when both $b\bar{b}$ are present.

Both D0 and CDF have searched for $3b$ final states present in the MSSM at
large $\tan\beta$. The D0 search involves a likelihood discriminant
which is optimized for the MSSM topology, and therefore cannot be applied to our $3b$ signal.
The CDF preliminary result in the $3b$ final state, using the invariant mass distribution $m_{12}$
of the two leading jets, has some excess of events consistent with
a new particle of mass in the $140-150$ GeV range and decaying to $b\bar{b}$.
The background normalizations are taken as free parameters, and are
fitted so that the deviation in $m_{12}$ is minimized. However, the shapes of
transverse energy distributions of the leading and second jets do not fit well
the standard model background \cite{CDF-bbb-url}. We have shown here that the ReCoM changes the shapes of these distributions such that they agree well
with the data (see Fig.~11). The best fit of the ReCoM to the CDF preliminary results indicates that the branching fraction of the scalar octet into two gluons is approximately 3 times larger than the branching ratio into  $b\bar{b}$.

Perhaps this good fit is only an accident, and the background modeling
performed by CDF can be modified such that the standard model fit to the
data improves. Nevertheless, the success of the ReCoM in describing the CDF data
is intriguing enough to warrant its experimental study.
Specifically, the CDF Collaboration could fit the background plus the ReCoM
signal to the $3b$ data for the $m_{12}$, $E_{T1}$ and $E_{T2}$ distributions.
The D0 Collaboration could use the scalar octet kinematics  to define the likelihood discriminant and to check whether the ReCoM is consistent with their observables.
Both collaborations could look for a $4b$ signal exhibiting a pair of resonances of
equal mass.  They could also search for the $bbjj$ signal from two resonances as a further test of the ReCoM.

The cross section for coloron production is large enough so that even the process where a $W$ boson is radiated 
from the initial state (see Fig. 12), which has smaller backgrounds, leads to a sufficient number of events to be tested at the Tevatron.
The signatures include a $W$ decaying leptonically and four jets, with the invariant mass distribution for two jets peaking at 
the same location as for the other two jets, near the scalar octet mass of $140 - 150$ GeV. When the two leading jets
come from the same scalar octet, the signature is a $W$ boson plus a dijet resonance, which may explain an excess in the 
CDF data from an inclusive search in this channel  \cite{CDF-WZ-url}. ReCoM predicts the presence of a second mass peak 
of the two leading jets, due to the direct decay of the coloron into quarks (see Fig. 14).

The ATLAS and CMS experiments can search for scalar octets with masses between a few hundred GeV and
a couple of TeV, impresively extending the Tevatron reach. In the case of a scalar octet with mass
of about $140-150$ GeV, relevant for the CDF excess, it is hard to overcome the
QCD background at the LHC, so that the Tevatron experiments should attempt to
discover or rule out its existence.

\bigskip\bigskip

{\bf Acknowledgments}:  We have benefited from insightful conversations on experimental aspects with Dan Amidei, 
Henry Frisch, Jonathan Hays and Tom Wright, and on theoretical aspects with 
Estia Eichten, Walter Giele, Roni Harnik, Patrick Fox, Gordan Krnjaic, Paul MacKenzie and Michael Peskin. SLAC is operated by Stanford University for the US Department of Energy under contract DE-AC02-76SF00515.
Fermilab is operated by Universities Research Association Inc.  under contract No. DE-AC02-76CH02000 with the DOE.

\vfil

\begin{thebibliography}{99} \frenchspacing

\bibitem{Hill:2002ap} For a review, see 
  C.~T.~Hill and E.~H.~Simmons,
  ``Strong dynamics and electroweak symmetry breaking,''
  Phys.\ Rept.\  {\bf 381}, 235 (2003)
  [Erratum-ibid.\  {\bf 390}, 553 (2004)]
  [arXiv:hep-ph/0203079].
  
\bibitem{Burdman:2006gy}
  G.~Burdman, B.~A.~Dobrescu and E.~Ponton,
  ``Resonances from two universal extra dimensions,''
  Phys.\ Rev.\  D {\bf 74}, 075008 (2006)
  [arXiv:hep-ph/0601186].

\bibitem{Kilic:2008pm}
  C.~Kilic, T.~Okui and R.~Sundrum,
  ``Colored Resonances at the Tevatron: Phenomenology and Discovery Potential
  in Multijets,''
  JHEP {\bf 0807}, 038 (2008)
  [arXiv:0802.2568 [hep-ph]].
``Vectorlike Confinement at the LHC,''
  JHEP {\bf 1002}, 018 (2010)
  [arXiv:0906.0577 [hep-ph]].

\bibitem{Bai:2010qg}
  Y.~Bai and R.~J.~Hill,
  ``Weakly interacting stable hidden sector pions,''
  Phys.\ Rev.\  D {\bf 82}, 111701 (2010)
  [arXiv:1005.0008 [hep-ph]].


\bibitem{Plehn:2008ae}
  T.~Plehn and T.~M.~P.~Tait,
  ``Seeking Sgluons,''
  J.\ Phys.\ G {\bf 36}, 075001 (2009)
  [arXiv:0810.3919 [hep-ph]].

\bibitem{Dobrescu:2007yp}
  B.~A.~Dobrescu, K.~Kong and R.~Mahbubani,
  ``Massive color-octet bosons and pairs of resonances at hadron colliders,''
  Phys.\ Lett.\  B {\bf 670}, 119 (2008)
  [arXiv:0709.2378].

\bibitem{Chivukula:1991zk}
  R.~S.~Chivukula, M.~Golden and E.~H.~Simmons,
  ``Multi - jet physics at hadron colliders,''
  Nucl.\ Phys.\  B {\bf 363}, 83 (1991).


\bibitem{Hill:1991at}
  C.~T.~Hill,
  ``Topcolor: Top quark condensation in a gauge extension of the standard model,''
  Phys.\ Lett.\  {\bf B266}, 419-424 (1991) ; \\
  C.~T.~Hill, S.~J.~Parke,
  ``Top production: Sensitivity to new physics,''
  Phys.\ Rev.\  {\bf D49}, 4454-4462 (1994).
  [hep-ph/9312324].

\bibitem{Chivukula:1996yr}
  R.~S.~Chivukula, A.~G.~Cohen, E.~H.~Simmons,
  ``New strong interactions at the Tevatron?,''
  Phys.\ Lett.\  {\bf B380}, 92-98 (1996).
  [hep-ph/9603311]; \\
  E.~H.~Simmons,
  ``Coloron phenomenology,''
  Phys.\ Rev.\  {\bf D55}, 1678-1683 (1997).
  [hep-ph/9608269].


\bibitem{nested} 
This term was first used in an early, unpublished version of Ref.~\cite{Dobrescu:2007yp}.


\bibitem{HiggsPlusb}
CDF Collaboration, ``Search for Higgs bosons produced in association with $b$ quarks'',
Note 10105, June 2010, \\
{\small \url{http://www-cdf.fnal.gov/physics/new/hdg//Results_files/results/3b_susyhiggs_jun10}}

\bibitem{CDF-bbb-url}
{\small \url{http://www-cdf.fnal.gov/physics/new/hdg//Results_files/results/3b_susyhiggs_jun10/more_plots.html}}

\bibitem{Manohar:2006ga}
  A.~V.~Manohar and M.~B.~Wise,
  ``Flavor changing neutral currents, an extended scalar sector, and the  Higgs
  production rate at the LHC,''
  Phys.\ Rev.\  D {\bf 74}, 035009 (2006)
  [arXiv:hep-ph/0606172]; \\
  M.~Gerbush, T.~J.~Khoo, D.~J.~Phalen, A.~Pierce and D.~Tucker-Smith, 
  ``Color-octet scalars at the CERN LHC,''
  Phys.\ Rev.\  {\bf D77}, 095003 (2008).
  [arXiv:0710.3133].

\bibitem{Gresham:2007ri}
  M.~I.~Gresham and M.~B.~Wise,
  ``Color Octet Scalar Production at the LHC,''
  Phys.\ Rev.\  D {\bf 76}, 075003 (2007)
  [arXiv:0706.0909 [hep-ph]].

\bibitem{Bai:2010mn}
  Y.~Bai and A.~Martin,
  ``Topological Pions,''
  Phys.\ Lett.\  B {\bf 693}, 292 (2010)
  [arXiv:1003.3006 [hep-ph]]. \\
  A.~R.~Zerwekh, C.~O.~Dib and R.~Rosenfeld,
  ``A New signature for color octet pseudoscalars at the CERN LHC,''
  Phys.\ Rev.\  D {\bf 77}, 097703 (2008)
  [arXiv:0802.4303].

\bibitem{Rizzo:1979mf}
  T.~G.~Rizzo,
  ``Gluon Final States In Higgs Boson Decay,''
  Phys.\ Rev.\  D {\bf 22}, 178 (1980)
  [Addendum-ibid.\  D {\bf 22}, 1824 (1980)].

\bibitem{Dobrescu:2010rh}
  B.~A.~Dobrescu, P.~J.~Fox and A.~Martin,
  ``CP violation in $B_s$ mixing from heavy Higgs exchange,''
  arXiv:1005.4238 [hep-ph].

\bibitem{Abazov:2010hv}
  V.~M.~Abazov {\it et al.}  [D0 Collaboration],
  ``Evidence for an anomalous like-sign dimuon charge asymmetry,''
  Phys.\ Rev.\  D {\bf 82}, 032001 (2010)
  [arXiv:1005.2757 [hep-ex]].

\bibitem{Becirevic:2001xt}
  D.~Becirevic, V.~Gimenez, G.~Martinelli, M.~Papinutto and J.~Reyes,
  ``B-parameters of the complete set of matrix elements of $\Delta B = 2$
  operators from the lattice,''
  JHEP {\bf 0204}, 025 (2002)
  [arXiv:hep-lat/0110091].

\bibitem{Gamiz:2009ku}
  E.~Gamiz, C.~T.~H.~Davies, G.~P.~Lepage, J.~Shigemitsu and M.~Wingate
                  [HPQCD Collaboration],
  ``Neutral $B$ Meson Mixing in Unquenched Lattice QCD,''
  Phys.\ Rev.\  D {\bf 80}, 014503 (2009)
  [arXiv:0902.1815 [hep-lat]].

\bibitem{Lenz:2006hd}
  A.~Lenz and U.~Nierste,
  ``Theoretical update of $B_s - \bar{B}_s$ mixing,''
  JHEP {\bf 0706}, 072 (2007)
  [arXiv:hep-ph/0612167].
    
\bibitem{Alwall:2007st}
  J.~Alwall {\it et al.},
  ``MadGraph/MadEvent v4: The New Web Generation,''
  JHEP {\bf 0709}, 028 (2007)
  [arXiv:0706.2334 [hep-ph]].

\bibitem{Pumplin:2002vw}
  J.~Pumplin, D.~R.~Stump, J.~Huston, H.~L.~Lai, P.~M.~Nadolsky and W.~K.~Tung,
  ``New generation of parton distributions with uncertainties from global QCD
  analysis,''
  JHEP {\bf 0207}, 012 (2002)
  [arXiv:hep-ph/0201195].

  \bibitem{pythia}
  T.~Sjostrand, S.~Mrenna and P.~Skands,
  ``PYTHIA 6.4 physics and manual,''
  JHEP {\bf 0605}, 026 (2006).
  [arXiv:hep-ph/0603175].

\bibitem{Abe:1998uz}
  F.~Abe {\it et al.}  [CDF Collaboration],
  ``Search for new particles decaying to $b\bar{b}$ in $p\bar{p}$ collisions at
  $\sqrt{s} = 1.8$ TeV,''
  Phys.\ Rev.\ Lett.\  {\bf 82}, 2038 (1999)
  [arXiv:hep-ex/9809022].


\bibitem{pgs}
J.~S.~Conway, ``Pretty Good Simulation of high-energy collisions'', 090401 release,
{\small \url{http://physics.ucdavis.edu/~conway/research/software/pgs/pgs4-general.htm} }

\bibitem{Abazov:2010ci}
  V.~M.~Abazov {\it et al.}  [D0 Collaboration],
  ``Search for neutral Higgs bosons in the multi-b-jet topology in 5.2 fb$^{-1}$ of
  $p\bar{p}$ collisions at $\sqrt{s} = 1.96$ TeV,''
  arXiv:1011.1931.



\bibitem{Cheung:2004ad}
  K.~Cheung and W.~Y.~Keung,
  ``Split supersymmetry, stable gluino, and gluinonium,''
  Phys.\ Rev.\  D {\bf 71}, 015015 (2005)
  [arXiv:hep-ph/0408335].\\
  Y.~Kats and M.~D.~Schwartz,
  ``Annihilation decays of bound states at the LHC,''
  JHEP {\bf 1004}, 016 (2010)
  [arXiv:0912.0526 [hep-ph]].\\
  C.~Kim and T.~Mehen,
  ``Color Octet Scalar Bound States at the LHC,''
  Phys.\ Rev.\  D {\bf 79}, 035011 (2009)
  [arXiv:0812.0307 [hep-ph]].

\bibitem{Martin:2009iq}
  A.~D.~Martin, W.~J.~Stirling, R.~S.~Thorne and G.~Watt,
  ``Parton distributions for the LHC,''
  Eur.\ Phys.\ J.\  C {\bf 63}, 189 (2009)
  [arXiv:0901.0002 [hep-ph]].

\bibitem{Djouadi:2005gj}
  For a review, see A.~Djouadi,
  ``The Anatomy of electro-weak symmetry breaking. II. The Higgs bosons in the
  minimal supersymmetric model,''
  Phys.\ Rept.\  {\bf 459}, 1 (2008)
  [arXiv:hep-ph/0503173].

\bibitem{Bardeen:1993ae}
  W.~A.~Bardeen and C.~T.~Hill,
  ``Chiral dynamics and heavy quark symmetry in a solvable toy field theoretic
  model,''
  Phys.\ Rev.\  D {\bf 49}, 409 (1994)
  [hep-ph/9304265]; \\ 
  W.~A.~Bardeen, E.~J.~Eichten and C.~T.~Hill,
  ``Chiral Multiplets of Heavy-Light Mesons,''
  Phys.\ Rev.\  D {\bf 68}, 054024 (2003)
  [arXiv:hep-ph/0305049].

\bibitem{Dobrescu:2009vz}
  B.~A.~Dobrescu, K.~Kong and R.~Mahbubani,
  ``Prospects for top-prime quark discovery at the Tevatron,''
  JHEP {\bf 0906}, 001 (2009)
  [arXiv:0902.0792 [hep-ph]].

\bibitem{Aaltonen:2008dn}
  T.~Aaltonen {\it et al.}  [CDF Collaboration],
  ``Search for new particles decaying into dijets in proton-antiproton
  collisions at $\sqrt{s} = 1.96$ TeV,''
  Phys.\ Rev.\  D {\bf 79}, 112002 (2009)
  [arXiv:0812.4036 [hep-ex]].

\bibitem{CDF-WZ-url}
{\small \url{http://www-cdf.fnal.gov/physics/ewk/2010/WW_WZ/index.html}}

\bibitem{Eichten:1996dx}
  E.~Eichten and K.~D.~Lane,
 ``Low - scale technicolor at the Tevatron,''
  Phys.\ Lett.\  B {\bf 388}, 803 (1996)
  [arXiv:hep-ph/9607213];
  ``Low-scale technicolor at the Tevatron and LHC,''
  Phys.\ Lett.\  B {\bf 669}, 235 (2008)
  [arXiv:0706.2339 [hep-ph]]. \\
  K.~Lane and A.~Martin,
  ``An Effective Lagrangian for Low-Scale Technicolor,''
  Phys.\ Rev.\  D {\bf 80}, 115001 (2009)
  [arXiv:0907.3737 [hep-ph]].


\end{thebibliography}
\end{document}